\documentclass[a4paper, 11pt]{article}
\usepackage{jcappub}
\usepackage{preamble}

\title{A Hamiltonian, post-Born, three-dimensional, on-the-fly
ray tracing algorithm for gravitational lensing}

\author[a]{Alan Junzhe Zhou,}
\author[b]{Yin Li,}
\author[a]{Scott Dodelson,}
\author[a]{Rachel Mandelbaum,}
\author[b]{Yucheng Zhang,}
\author[a]{Xiangchong Li,}
\author[c]{\& Giulio Fabbian}

\affiliation[a]{Department of Physics, Carnegie Mellon University, Pittsburgh, PA 15213, USA\\
                McWilliams Center for Cosmology and Astrophysics, Carnegie Mellon University, Pittsburgh, PA 15213}
\affiliation[b]{Department of Mathematics and Theory, Peng Cheng Laboratory, Shenzhen, Guangdong 518000, China}
\affiliation[c]{Institute of Astronomy, Madingley Road, Cambridge CB3 0HA, UK\\
                Kavli Institute for Cosmology Cambridge, Madingley Road, Cambridge CB3 0HA, UK\\
                School of Physics and Astronomy, Cardiff University, The Parade, Cardiff, CF24 3AA, UK\\
                Center for Computational Astrophysics, Flatiron Institute, New York, New York 10010, USA}

\emailAdd{ajzhou@uchicago.edu}
\emailAdd{eelregit@gmail.com}

\abstract{The analyses of the next generation cosmological surveys
demand an accurate, efficient, and differentiable method for simulating
the universe and its observables across cosmological volumes.
We present Hamiltonian ray tracing (HRT) --- the first post-Born
(accounting for lens-lens coupling and without relying on the Born approximation),
three-dimensional (without assuming the thin-lens approximation),
and on-the-fly (applicable to any structure formation simulations)
ray tracing algorithm based on the Hamiltonian formalism.
HRT performs symplectic integration of the photon geodesics in a weak
gravitational field, and can integrate tightly with any gravity solver,
enabling co-evolution of matter particles and light rays with minimal
additional computations.
We implement HRT in the particle-mesh library \pmwd, leveraging hardware
accelerators such as GPUs and automatic differentiation capabilities based on \jax.
When tested on a point-mass lens, HRT achieves sub-percent accuracy in
deflection angles above the resolution limit across both weak and
moderately strong lensing regimes.
We also test HRT in cosmological simulations on the convergence maps and
their power spectra.
}

\begin{document}
\maketitle
\flushbottom

\section{Introduction}
\label{sec:introduction}

Next-generation galaxy, lensing, and Cosmic Microwave Background (CMB)
surveys promise percent-level or better constraints on cosmological and
astrophysical parameters.
This discovery potential, however, hinges on our ability to quickly and
accurately simulate the universe and its observables over cosmological
volumes.
For example, computing covariance matrices for two-point cosmological
analyses using mock catalogs requires a large number of mock catalogs
\cite{KiesslingEtAl2011, FosalbaEtAl2015, SgierEtAl2019, TessoreEtAl2023}.
Higher-order statistical methods, such as peak and void counts, require
many high-fidelity simulations to model how observables and their
covariances depend on cosmology \cite{li_constraining_2019,
liu_cosmology_2015}.
Simulation-based inference (SBI) needs a large number of simulations for
its training set \cite{cranmer_frontier_2020}.
Finally, field-level inference (FLI) demands a fast and differentiable
forward model to simulate the universe and to construct observables
thereof \cite{LiEtAl2024, pmwd, zhou_field-level_2023,
zhou_accurate_2023, BORG, ELUCID, SeljakEtAl2017, BORG-PM,
SchmidtEtAl2019, alsing_cosmological_2017, anderes_bayesian_2015,
KitauraEtAl2021, AtaEtAl2021, fiedorowicz_karmma_2022,
fiedorowicz_karmma_2022-1, millea_optimal_2021,
lanzieri_forecasting_2023, porqueres_field-level_2023, NguyenEtAl2024}.
Perhaps not surprisingly, photons are the primary carriers of
information for most cosmological observables.
A photon conveys information from both its source (e.g., the temperature
or polarization of the primordial CMB, or the positions of galaxies) as
well as the intervening matter structures through which it traverses.
In this work, we propose a new ray tracing algorithm called Hamiltonian
ray tracing (HRT) that 1) accounts for the actual light trajectory and
lens-lens couplings (post-Born effects in short); 2) is differentiable
and computed on-the-fly; and 3) incurs minimal computational overhead
when integrated with a background N-body algorithm.

The Born approximation assumes that lensing observables (e.g., weak
lensing convergence $\kappa$ or the CMB deflection angle) accumulate
along a photon's unperturbed path and consider subsequent lensing events
as independent from each other.
Post-Born corrections effectively accounts for the ray deflection during
a photons propagation and for the so-called lens-lens coupling, which
describes how gravitational lenses at different redshifts can interact
to generate rotational modes in the observable fields
\cite{dodelson_second_2005, cooray_second_2002, krause_hirata_2010}.
Post-Born effects introduce additional non-Gaussianities into lensing
observables besides those generated by the non-linear clustering of the
matter.
Thus, it is important to include them in the modeling of higher-order
statistics for both CMB and galaxy lensing.
In particular, in the case of CMB lensing, non-linear and post-Born
effects have similar amplitude \cite{pratten_impact_2016,
fabbian_et_al_2018}.
For example, post-Born effects significantly change the higher-order
moments or peaks statistics of galaxy weak lensing convergence maps
\cite{petri_validity_2017, fabbian_et_al_2019}.
These effects translate into cosmological biases for analysis that
targets the aforementioned higher-order statistics.
Moreover, post-Born corrections affect FLI through the galaxy delensing
effect.
As demonstrated by Ref.~\cite{chang_delensing_2014}, galaxies at $z=1$
are typically deflected on an arc minute scale, which is the typical
pixel size in an FLI analysis.
If the post-Born effects are unaccounted for, the FLI's forward model
will generate observable fields that are coherently shifted with respect
to the truth.
However, since all existing lensing FLI models to date assume the Born
approximation \cite{porqueres_field-level_2023,
lanzieri_forecasting_2023, bohm_madlens_2020} and there does not yet
exist a forward model that captures post-Born effects, the resulting
cosmological biases are not well understood.
Post-Born effects are less disruptive for two-point statistics, although
the corrections could still be measured for the highest multipoles of the
shear and lensing convergence power spectra at the level of sensitivity of future
CMB lensing experiments and galaxy surveys \cite{cooray_second_2002, dodelson_second_2005,
dodelson_reduced_2006, jain_raytracing_2000, hirata_reconstruction_2003,
hilbert_accuracy_2020, beck_et_al_2018, bohm_et_al_2018, fabbian_et_al_2019}. Similar conclusions apply for cross-correlations between
CMB lensing and galaxy survey probes \cite{bohm_et_al_2020}.

Weak lensing simulations typically capture post-Born effects using the
multiple lens plane (MLP) formalism \cite{vale_simulating_2003,
jain_raytracing_2000, hilbert_strong_2007, hilbert_ray-tracing_2009},
which has been widely applied in many studies
\cite{sato_simulations_2009, Becker2013, petri_mocking_2016,
takahashi_full-sky_2017, osato_kappatng_2021, petri_et_al_2017,
bohm_et_al_2018, fabbian_et_al_2018, WeiEtAl2018, XuJing2021,
Jimenez-VicenteMediavilla2022, SuoEtAl2023}
Our new HRT formalism improves upon the MLP method in ways that enable
new cosmological analysis methods.
\textbf{First}, whereas MLP deflects photons only by the gravitational
potential gradients instantaneously at a lens plane, HRT utilizes the
gravitational potential of the entire simulated volume without the
thin-lens approximation, a feature known as three-dimensional ray
tracing \cite{CouchmanEtAl1999, vale_simulating_2003, LiEtAl2011,
barreira_ray-ramses_2016, KilledarEtAl2012}.
\textbf{Second}, the MLP algorithm requires initially running an N-body
simulation, storing its time steps (called snapshots), preparing the
density field on the light cone in mass shells, and finally accumulating
the convergence fields backward in time.
This multi-step process decouples ray tracing from the N-body
simulations, complicating implementation when varying cosmology and
introducing computational overhead in terms of data storage and
transfer.
Our HRT method performs ray tracing on-the-fly by co-evolving the light
rays with the matter particles.
The physics and the time-stepping of the rays are tightly synchronized
with the N-body simulation and subject to the same cosmological model,
thereby eliminating the need for data storage.
Refs.~\cite{LiEtAl2011, barreira_ray-ramses_2016}'s algorithm performs
three-dimensional ray tracing on-the-fly, but ignore post-Born effects
and lack differentiability and other features listed below.
\textbf{Third}, FLI requires that the observable fields be
differentiable with respect to all cosmological parameters and the
initial conditions of the background N-body simulation.
Partly because of the multi-step process, current MLP codes are not
differentiable.
HRT formalism leverages a combination of automatic differentiation
routines in \jax\ and variational equation to achieve efficient
differentiation with respect to all parameters of interest.
\textbf{Fourth}, MLP typically uses Kaiser-Squires algorithm
\cite{jain_raytracing_2000} or finite-difference
\cite{hilbert_ray-tracing_2009} methods to compute lensing observables,
which could suffer from numerical instability.
As we shall see, since most lensing observables are defined in terms of
derivatives of the photon trajectory with respect to its initial
position, and our new methods are differentiable, we can calculate these
observables by direct differentiation.
\textbf{Fifth}, unlike MLP which approximates the total lensing deflection as a Riemann sum over the radial comoving space, HRT solves each photon's (light ray's)  equations of motion (EOMs)
using a highly stable and accurate symplectic integrator. This
provides a systematic way to correct for the finite time-resolution
of the snapshots. 
\textbf{Finally}, MLP implementations do not currently leverage hardware
accelerators while HRT is implemented on both CPUs and GPUs. \response{Moreover, on-the-fly ray-tracing is particularly suited for GPU applications since GPU-based N-body simulations are already memory limited and cannot afford to store snapshots and post-process them later into shear maps.}

The paper is organized in the following way.
We first review the Hamiltonian formulation of the photon dynamics in a
gravitational field in \cref{sec:hamiltonian}.
Next, we discretize the EOMs to construct a symplectic integrator for
the photons' trajectories in \cref{sec:symplectic}.
Once we have the trajectories, we construct an algorithm that calculates
the weak lensing observables along the light path in
\cref{sec:observables}.
We discuss an efficient implementation of the algorithm in
\cref{sec:implementation}, and test the algorithm for point mass
deflection and weak lensing maps in cosmological volumes in
\cref{sec:validations}.
We conclude in \cref{sec:conclusion}.

\section{Hamiltonian dynamics of light rays}
\label{sec:hamiltonian}

In post-Born ray tracing, the trajectories of photons are deflected by
the large-scale structure.
Consequently, it is not known a priori if a photon emitted by a source
will reach the observer.
The core idea of HRT is to evolve both light rays (from its observed
position on the image plane, denoted by $\bftheta_0$) and the matter
particles (from their displacements and velocities at today) backward in
time in an N-body simulation\footnote{We can backward evolve the dark
matter particles by reversing the time variable in their EOMs. \response{Table 2 of Ref.~\cite{LiEtAl2024} has shown that both the positions and velocities of the dark matter particles can be recovered to high accuracy across different simulation configurations in \pmwd{} when the EOMs are solved with a symplectic integrator.}}
The light ray EOMs are reversible once we describe them with Hamiltonian
dynamics, which we then implement numerically using symplectic
integration.
This approach involves three parts: constructing the spacetime metric
with gravitational perturbations, defining the photon's Hamiltonian
using this metric, and finally, solving Hamilton's equations.
We only motivate the results in this section and leave the detailed
derivations to \cref{app:hamiltonian}.

Let us parameterize spacetime with conformal time $\tau$ and comoving
spatial coordinates $\bfx$.
On scales smaller than the cosmic horizon, the matter density contrast
$\delta(\bfx, \tau)$ sources a gravitational potential field
$\npot(\bfx, \tau)$ via Poisson's equation,
\begin{align}
    \grad^2 \npot(\bfx, \tau)
     & =
    \frac{3}{2}
    \frac{\Omegam H_0^2}{a(\tau)}
    \delta(\bfx,\tau) \,,
\end{align}
where the gradient (unless specified) is with respect to the comoving
coordinates.
Here, $a$, $\Omegam$, and $H_0$ represent the scale factor, matter
fraction, and Hubble parameter, respectively.
The potential $\npot(\bfx, \tau)$ perturbs the otherwise homogeneous and
isotropic Friedmann-Lema\^itre-Robertson-Walker metric, given by
\begin{align}
    \label{eqn:metric_4d}
    ds^2 &=
    - a^2 \Bigl( 1 + 2\frac{\npot}{c^2} \Bigr) d\tau^2
    + a^2 h_{ij} dx^i dx^j \\
    \label{eqn:metric_3d}
    &\equiv
    - a^2 \Bigl( 1 + 2\frac{\npot}{c^2} \Bigr) d\tau^2
    + a^2 \Bigl( 1 - 2\frac{\npot}{c^2} \Bigr) (r^2 d\bftheta^2 + d\chi^2) \,,
\end{align}
for $i = 1, 2, 3$.
$r$ and $\chi$ are the transverse and radial comoving distances,
respectively, and $r = \chi$ in a flat universe.
We parameterize the sky under the flat-sky approximation with angular
coordinates, i.e., $\bfx = (\bftheta, \chi)$.

Now, consider a photon.
The general relativistic covariant Hamiltonian of a photon is given by
\begin{equation}
\label{eqn:hamiltonian_3d}
H(x^j, p_i, \tau) = c \sqrt{h^{ij}p_i p_j}
  \Bigl(1 + 2\frac{\npot}{c^2}\Bigr) \,,
\end{equation}
where $p_i$ is the momentum one-form \cite{bar-kana_gravitational_1997}.
The EOMs of the photon are solutions to Hamilton's equations.
The results are (see detailed proof and discussion in
\cref{app:hamiltonian}; Ref.~\cite{bar-kana_gravitational_1997} also
derives a similar result albeit with a different parameterization of the
image plane)
\begin{align}
    \label{eqn:eom1}
    \frac{d\bftheta}{d\chi}
     & = - \frac{\bfeta}{r^2} \,,                                 \\
    \label{eqn:eom2}
    \frac{d\bfeta} {d\chi}
     & = \frac{2}{c^2} \frac{\partial\npot}{\partial \bftheta} \,,\\
    \label{eqn:eom3}
    \frac{d\tau}{d\chi}
     & = - \frac{1}{c}\left(1 - 2\frac{\npot}{c^2} + \frac{\bfeta^2}{2r^2} \right)\,,
\end{align}
where we have defined the conjugate momentum
\begin{equation}
\bfeta = \frac{r \bfv}{c} \,,
\end{equation}
and $\bfv$ is the \emph{transverse} peculiar velocity of the photon.
\Cref{eqn:eom1} is simply a geometric relation in the deflection
tangent: $- r d\bftheta / d\chi = \bfv / c$.
The second and third terms in \cref{eqn:eom3} correspond to the Shapiro and the geometric time delay, respectively.
$\bftheta$ has the unit of angle while $\bfeta$ has the unit of
length\footnote{They are actually conjugate variables in a separable
Hamiltonian and relates to the Fermat action as shown in
\cref{app:symp}.
In fact, the time delay terms can also be seen easily from the separable
Hamiltonian.}.
The initial conditions are given by
\begin{equation}
    \bftheta(\chi_0) = \bftheta_0 \,, \quad
    \bfeta(\chi_0) = 0 \,, \quad
    \chi_0 = 0 \,.
\end{equation}
To arrive at these EOMs, we have used the flat-sky (small angle)
approximation, kept $\npot/c^2$, $\grad\npot/c^2$, and \response{assumed $(\bfv/c)^2$ is of $\order(\npot/c^2$).}
We have also used the comoving distances instead of conformal time to
parameterize the photon trajectory.
The two are related by \cref{eqn:eom3}, where the second term is the
Shapiro time delay.
However, we will not consider the time delay correction in our study.

\section{Symplectic ray tracing}
\label{sec:symplectic}

\subsection{Equations of motion in discrete time}

So far, we have derived the EOMs for photons in angular coordinates and
conjugate momenta using the Hamiltonian formalism.
We now introduce a second-order symplectic integrator to integrate these
EOMs and trace the photons' trajectories.
A symplectic integrator advances the equations of motion in discrete
time steps, each of which preserves the mathematical structure of the
Hamiltonian.
As such, a symplectic integrator maintains the long-term behavior of the
particle's motion.
Symplectic integrators have been widely used in cosmological simulations
and achieve superior accuracy compared to other integration methods of
the same or even higher orders \cite{quinn_time_1997,
saha_symplectic_1992, springel_simulating_2021}.

In this work, we utilize the kick-drift-kick (KDK) integrator, also
known as the velocity Verlet method \cite{yoshida_recent_1993}.
We will only sketch out the key ideas here, and leave a more detailed
analysis of the integrator to \cref{app:symp}.
The KDK integration scheme operates by decomposing each integration step
of the EOMs, $\Delta \chi$, into three consecutive stages, updating
either the positions or the momenta at a time.
Initially, we update $\bfeta$ for a half time step ($\Delta \chi / 2$),
leaving $\bftheta$ unchanged (the first kick).
Next, we update $\bftheta$ for a full time step ($\Delta \chi$), while
keeping $\bfeta$ unchanged (the drift stage).
Finally, we apply another half time step update to $\bfeta$ to complete
the integration (the second kick).

To implement this integration scheme, we configure the time steps in our
simulations as follows.
We perform ray tracing while evolving the N-body simulation backward in
time.
Light is traced from the observer's position at $\chi = 0$ and $\tau =
\tau_0$, towards the light source at $\chi_s > 0$ and $\tau = \tau_0 -
\chi_s/c$.
We define a series of lens in mass shells perpendicular to the main line
of sight, labeled by the subscript $n$, with $n=0$ at the observer and
$n=s$ at the source.
The lens shells are determined by the time stepping of the forward
simulation itself, with each shell sandwiched between the light fronts
at two consecutive time steps, and the $n$-th shell's midpoint labeled
by $n+1/2$.
With this set up, we can now integrate the EOMs iteratively via (see
\cref{app:symp} for derivation and error analysis)
\begin{align}
\label{eqn:symp_1}
\bfeta_{n+\half}
&= \bfeta_n + K_n^{n+1/2}(\bftheta_n, \chi_n) \,, \\
\label{eqn:symp_2}
\bftheta_{n+1}
&= \bftheta_n + D_n^{{n+1}}(\bfeta_{n+1/2}, \chi_{n+1/2}) \,, \\
\label{eqn:symp_3}
\bfeta_{n+1}
&= \bfeta_{n+1/2} + K_{n+1/2}^{n+1}(\bftheta_{n+1}, \chi_{n+1}) \,,
\end{align}
where
\begin{align}
\label{eqn:kick_notimecor}
K_{a}^{b}(\bftheta_c, \chi_c)
&= - \frac{2}{c} \int_{\chi_a}^{\chi_b} d\chi \,
  r(\chi) \grad_\perp \npot (\bfx_c, \tau(\chi)) \,, \\
\label{eqn:drift}
D_{a}^{b}(\bfeta_c, \chi_c)
&\simeq (\frac{\bfeta_c}{c r_c^2})(\chi_b-\chi_a) \,,
\end{align}
are the kick and the drift factors, and $\grad_\perp \equiv r^{-1}
\grad_\bftheta$ denotes the gradient with respect to the transverse
comoving coordinates.
Note that the traditional velocity Verlet algorithm typically
approximates the first kick operator in \Cref{eqn:symp_1} as $\frac12
K_{n-1/2}^{n+1/2}(\bftheta_n, \chi_n)$.
The proposed iteration thus provides a better comoving space (time)
resolution.

\subsection{Time resolution correction}

\Cref{eqn:kick_notimecor} updates light ray's conjugate momenta by
integrating the \emph{time-evolving} $\grad_\perp \npot (\bfx,
\tau(\chi))$ along their trajectories.
In practice, we derive $\grad_\perp \npot (\bfx, \tau(\chi))$ from the
forces in N-body simulation, at discrete time steps often referred to as
"snapshots", say at $\tau(\chi_c)$.
If each time step is small, we can approximate the growth of the matter
structure using
\begin{equation}
\npot (\bfx, \tau(\chi)) = \npot (\bfx, \tau_c)
  \frac{D_1(\tau)}{D_1(\tau_c)} \frac{a(\tau_c)}{a(\tau)}\,,
\end{equation}
where $D_1$ is the linear growth function, $\tau = \tau(\chi)$, and
$\tau_c = \tau(\chi_c)$.
Using this correction, \cref{eqn:kick_notimecor} becomes
\begin{equation}
\label{eqn:kick}
K_{a}^{b}(\bftheta_c, \chi_c)
= - \frac{2}{c} \int_{\chi_a}^{\chi_b} d\chi \,
  r(\chi) \grad_\perp \npot (\bfx_c, \tau)
  \frac{D_1(\tau)}{D_1(\tau_c)} \frac{a(\tau_c)}{a(\tau)} \,.
\end{equation}

\section{Weak lensing observables}
\label{sec:observables}

In \cref{sec:symplectic}, we worked out the trajectories of the photons.
We now aim to construct observable maps on the image plane using these
photon trajectories.
Most cosmological observables $O$ can be written as line-of-sight
integrals along the light path,
\begin{equation}
\label{eq:integral}
O(\bftheta) = \int_{\bftheta(\chi)} d\chi \,
  P(\chi) Q(\bfx,\tau(\chi))\,,
\end{equation}
where $P$ is the line-of-sight kernel, and $Q$ is any cosmological
field.
Examples include weak lensing distortions, the galaxy density, the
Sunyaev-Zel’dovich effects, the integrated Sachs-Wolfe effect, and the
dispersion measure \cite{barreira_ray-ramses_2016, dodelson_modern_2003,
schneider_gravitational_1992}.
Here, we focus on weak lensing and develop an efficient and accurate
algorithm for computing the convergence, shear, and rotation maps.

We begin with the distortion matrix $\bfA$, which characterizes the
lensing effect of an object at source $\chi$,
\begin{align}
\label{eqn:A}
\bfA(\chi) = \frac{\partial\bftheta(\chi)}{\partial\bftheta_0} \,.
\end{align}
$\bfA$ can be decomposed into the product of a rotation and a shear
matrix\footnote{omitting $\chi$ for clarity}:
\begin{equation}
\bfA = \begin{pmatrix}
  \cos \omega  & \sin \omega \\
  -\sin \omega & \cos \omega
\end{pmatrix}
\begin{pmatrix}
  1-\kappa-\gamma_1 & -\gamma_2 \\
  -\gamma_2         & 1-\kappa+\gamma_1
\end{pmatrix} \,.
\end{equation}
The convergence $\kappa$, shear $\gamma$, and rotation $\omega$ are the
primary weak lensing observables.
Since $\omega$ is orders of magnitude smaller than $\kappa$ and
$\gamma$, we solve for the observables in terms of the entries of $\bfA$
keeping $\omega$ to first order,
\begin{equation}
\left\{\begin{aligned}
  \kappa &= 1 - C_1 \,, &\quad
  \gamma_1 &= C_3 + C_4 \, \omega \,, \\
  \omega &= \frac{C_2}{1-\kappa} \,, &\quad
  \gamma_2 &= - C_4 + C_3 \, \omega \,, \\
\end{aligned}\right.
\label{eqn:obs_first_set}
\end{equation}
where
\begin{equation}
\left\{\begin{aligned}
  C_1 &= \frac{A_{11} + A_{22}}{2} \,, &\quad
  C_3 &= \frac{A_{22} - A_{11}}{2} \,, \\
  C_2 &= \frac{A_{12} - A_{21}}{2} \,, &\quad
  C_4 &= \frac{A_{12} + A_{21}}{2} \,.
\end{aligned}\right.
\label{eqn:obs_second_set}
\end{equation}
Finally, going back to \cref{eq:integral}, the line-of-sight kernel $P$
is the normalized galaxy density distribution $n_\mathrm{g}(\chi)$,
and $Q$ is either $\kappa$, $\gamma$, or $\omega$.

Past works have computed these observables using either the
Kaiser-Squires inversion \cite{jain_raytracing_2000} or finite
differences \cite{barreira_ray-ramses_2016, hilbert_ray-tracing_2009}.
However, the former is sensitive to boundary conditions and the latter
can induce numerical instability.
We propose a new algorithm that computes the observable maps via direct
(forward-mode) differentiation of \cref{eqn:A}, which is convenient to
implement in our framework using \jax\ automatic differention.

The main idea is to decompose $A_n$\footnote{We use the subscript $n$
to denote observables computed at $\chi_n$, e.g., $A(\chi_n)=A_n$.} as
chained products of Jacobian matrices for time steps $n' < n$.
Differentiating the discretized EOM in \cref{eqn:symp_2} with respect to
the initial ray position using the chain rule,
\begin{align}
\nonumber
\frac{\partial\bftheta_{n+1}}{\partial\bftheta_0}
&= \frac{\partial \bftheta_n}{\partial \bftheta_0}
  + \frac{\partial D_n^{n+1}}{\partial \bfeta_{n+1/2}}
  \frac{\partial \bfeta_{n+1/2}}{\partial \bftheta_0} \\
\label{eqn:obsv}
&= \frac{\partial \bftheta_n}{\partial \bftheta_0}
  + \frac{\partial D_n^{n+1}}{\partial \bfeta_{n+1/2}}
    \biggl( \frac{\partial \bfeta_n}{\partial \bftheta_0}
      + \frac{\partial K_n^{n+1/2}}{\partial \bftheta_n}
      \frac{\partial \bftheta_n}{\partial \bftheta_0} \biggr)
\end{align}
Using the helper function
\begin{equation}
\label{eqn:B}
\bfB(\chi) = \frac{\partial\bfeta(\chi)}{\partial\bftheta_0} \,,
\end{equation}
we can then rewrite \cref{eqn:obsv} as an iteration,
\begin{align}
\label{eqn:obsv_recurse1}
\bfB_{n+1/2} &= \bfB_n
  + \frac{\partial K_n^{n+1/2}}{\partial \bftheta_n} \bfA_n \,, \\
\label{eqn:obsv_recurse2}
\bfA_{n+1} &= \bfA_n
  + \frac{\partial D_n^{n+1}}{\partial \bfeta_{n+1/2}} \bfB_{n+1/2} \,, \\
\label{eqn:obsv_recurse3}
\bfB_{n+1} &= \bfB_{n+1/2}
  + \frac{\partial K_{n+1/2}^{n+1}}{\partial \bftheta_{n+1}} \bfA_{n+1} \,,
\end{align}
with the initial conditions (for each photon),
\begin{equation}
\bfA_{0} = \mathbb{I}_{2\times2} \,, \quad
\bfB_{0} = \mathbb{O}_{2\times2} \,.
\end{equation}
Here, $\mathbb{I}$ and $\mathbb{O}$ are the identity and the zero
matrices, respectively.
We compute the iterations for $\bfA$ and $\bfB$ while calculating the
trajectories of the photons.
Each $\bfA$ and $\bfB$ iteration step is a Jacobian-vector product,
allowing us to calculate them automatically using the forward
differentiation in \jax, with minimal computational and memory overhead.
Finally, we accumulate the observables by
\begin{equation}
\label{eqn:obs_accum}
O = \sum_n n_\mathrm{g}(\chi_n) \, O_n \,,
\end{equation}
where $O_n \in \{\kappa_n, \gamma_n, \omega_n\}$.

\section{Implementation}
\label{sec:implementation}

\subsection{Efficient dark matter and photons co-evolution}

\begin{algorithm}[t]
\caption{Reverse-time co-evolution of dark matter and a single ray}
\label{alg:ray_tracing}
\begin{algorithmic}
  \Procedure{raytracing}{$\photon\,, \dm\,, n_g$}
  \State $A \gets \mathbb{I}_{2\times2} \,, B \gets \mathbb{O}_{2\times2}$
  \State $[\kappa\,, \gamma_1\,, \gamma_2\,, \omega] \gets \mathbf{0}$
  \State $a \gets 1$
  \While{$a \geq a_\mathrm{min}$} \Comment{ray-trace backward in time to $a_\mathrm{min}$}
    \State $\grad_\perp \npot \gets \dm$
    \State $\dm \gets \texttt{nbody\_reverse\_step}(\dm\,, \grad_\perp \npot)$
      \Comment{Evolve dark matter backward \cite{LiEtAl2024}}
    \State $\photon \gets \texttt{kick}(\photon\,, \grad_\perp \npot)$
      \Comment{\cref{eqn:eom1}}
    \State $B \gets \texttt{iterate\_B}(\photon\,, A\,, B\,, \grad_\perp \npot)$
      \Comment{\cref{eqn:obsv_recurse1}}
    \State $\photon \gets \texttt{drift}(\photon)$
      \Comment{\cref{eqn:eom2}}
    \State $A \gets \texttt{iterate\_A}(\photon\,, A\,, B)$
      \Comment{\cref{eqn:obsv_recurse2}}
    \State $\photon \gets \texttt{kick}(\photon\,, \grad_\perp \npot)$
      \Comment{\cref{eqn:eom1}}
    \State $B \gets \texttt{iterate\_B}(\photon\,, A\,, B\,, \grad_\perp \npot)$
      \Comment{\cref{eqn:obsv_recurse3}}
    \State $\kappa\,, \gamma_1\,, \gamma_2\,, \omega \gets \texttt{observe}(A\,, n_g\,, \kappa\,, \gamma_1\,, \gamma_2\,, \omega)$
      \Comment{\cref{eqn:obs_first_set,eqn:obs_accum}}
    \State $a \gets a - \Delta a$
  \EndWhile
  \State \Return $\kappa\,, \gamma_1\,, \gamma_2\,, \omega$
  \EndProcedure
\end{algorithmic}
\end{algorithm}

In this section, we propose an implementation of the HRT algorithm that
co-evolves the dark matter particles and light rays\footnote{Each ray
(bundle) is defined as the collection of photons observed in the same
pixel on the image plane.}.
This implementation only incurs minimal computational overhead compared
to evolving the dark matter particles alone.
The HRT algorithm is applicable to all numerical simulations of
structure formation, and particularly so to differentiable ones.
Specifically, we have implemented it atop the \pmwd\ library, which
offers a differentiable, fast, and memory-efficient particle mesh-based
dark matter simulation \cite{LiEtAl2024, pmwd}.

Let us first evolve a dark matter distribution from the initial
condition to $a=1$ where the observer is.
To start ray tracing, we first initialize the rays on a uniform grid
representing the pixels (of size $\mu_\mathrm{2D}$; \cref{tab:variables}
lists the mesh variables in this section) on the image plane at $a=1$.
To simplify our discussion, we will focus on tracing a single ray.
The actual implementation parallelizes trivially across all the rays
since there are no interactions between them.
We use $\photon$ to denote the state (position and momentum) of this ray
and $\dm$ to represent the state of all the dark matter particles.
We co-evolve the matter particles and the light rays backward in time as
described in \cref{alg:ray_tracing}.
At each time step, we compute the potential gradient $\grad_\perp \npot$
on the 3D mesh.
We then use $\grad_\perp \npot$ to first evolve the dark matter
particles ($\texttt{nbody\_reverse\_step}$ in \cref{alg:ray_tracing},
\cite{LiEtAl2024}) backward in time and then to integrate the ray's EOMs
via the KDK integrator (\cref{sec:symplectic}).

\begin{table}[htbp]
  \centering
  \caption{Definitions of mesh and ray variables.
  The white rows define the resolution of meshes or rays, while the
  light gray ones define their corresponding position vectors.
  The top two rows are for the 3D particle mesh, on which the
  gravitational forces are computed.
  The middle two rows specify the 2D configuration of rays.
  And the bottom two rows describe the 2D ray mesh on which we
  interpolate and transfer the transverse gravitational forces, from the
  3D particle mesh and to the 2D ray positions.}
  \label{tab:variables}
  \begin{tabular}{@{}cl@{}}
    \tblhc
    \textbf{Variable} & \textbf{Purpose / Definition} \\
    $\resthreed$ & 3D particle mesh comoving resolution \\
    \tblrc
    $\xthreed$ & 3D positions of particle mesh grid points \\
    $\pixsize$ & 2D ray spacing / pixel resolution on the image plane \\
    \tblrc
    $\bftheta_0, \bftheta$ & 2D ray positions on the image plane and
                             during ray tracing, respectively \\
    $\restwod$ & 2D ray mesh resolution during ray tracing \\
    \tblrc
    $\xtwod$ & 2D positions of ray mesh grid points during ray tracing \\
  \end{tabular}
\end{table}

The most computationally expensive operation in the simulation is to
calculate $\grad_\perp \npot$ via the Fast Fourier Transform (FFT)
\cite{LiEtAl2024}.
\Cref{alg:ray_tracing}, however, only requires $\grad_\perp \npot$ to be
calculated once per time step.
The extra computation is either in 2D or on a thin shell of 3D field
(e.g., the computation of the kick operator as detailed below in
\cref{sec:kick}), and thus is negligible compared to the 3D work load
already done by the gravity solvers.

\subsection{The kick operator}
\label{sec:kick}

In a particle mesh (PM)-based N-body simulation, we first compute the
gravitational force field on a 3D mesh and then interpolate this force
onto particle positions.
Similarly, we compute the force on light rays (in the kick operator,
$K_a^b(\bftheta_c, \chi_{c})$ in \cref{eqn:kick}, where $\bftheta_c$ is
the position of the ray at the integration step) using the PM method.
Our algorithm achieves this using 3 meshes/arrays: a 3D particle mesh
covering the entire simulation volume with resolution $\resthreed$ and
coordinates $\xthreed$, a 2D array of rays covering the image plane with
angular resolution $\pixsize$ and coordinates $\bftheta_0$, and an
intermediate 2D ray mesh transferring forces from the first mesh to the
second with angular resolution $\restwod$ and coordinates $\xtwod$.

We first evaluate the integrand of \cref{eqn:kick} on the 3D mesh.
For mesh points within the lens plane (having comoving coordinates
$\chi_a \leq \chi < \chi_b$), we calculate their projected angles, then
interpolate and accumulate their values onto the ray mesh using
cloud-in-cell (CIC), or trilinear, interpolation
\cite{hockney_computer_2021}.
This process evaluates the kick operator line-of-sight integration on
the ray mesh, $K_a^b(\xtwod, \chi_c)$.
To account for the smoothing effect introduced by the CIC interpolation,
we deconvolve $K_a^b(\xtwod, \chi_c)$ in Fourier space as follows,
\begin{equation}
    \fs K_a^b(\bfk, \chi_c) \to \fs K_a^b(\bfk, \chi_c) ~
    \mathrm{sinc}\Bigl( \frac{\bfk_x \restwod}2 \Bigr)^{-4} ~
    \mathrm{sinc}\Bigl( \frac{\bfk_y \restwod}2 \Bigr)^{-4} \,,
\end{equation}
where $\fs\cdot$ denotes the Fourier space quantity and $\bfk$ is the
wave vector.
And the powers of -4 on the sinc functions accounts for both the
interpolation from the 3D particle mesh to the 2D ray mesh, and that
from the latter to the rays.
To account for the finite width of the ray bundle, we apply Gaussian
smoothing on $K_a^b(\xtwod, \chi_c)$ at the resolution
\begin{equation}
\label{eqn:reslim}
\reslim = \max\Bigl( \frac{2 \, \resthreed}{r_a + r_b}, \pixsize \Bigr) \,,
\end{equation}
by applying
\begin{equation}
\fs K_a^b(\bfk, \chi_c) \to \fs K_a^b(\bfk, \chi_c)
\exp\Bigl( \frac12 (\bfk_x^2+\bfk_y^2) \reslim^2 \Bigr) \,.
\end{equation}
Finally, we use CIC interpolation to map $K_a^b(\xtwod, \chi_c)$ onto
the positions of the rays, resulting in $K_a^b(\bftheta_c, \chi_c)$.

\subsection{Adaptive ray mesh}
\label{sec:ray_mesh}

Unlike the 3D mesh which has a fixed comoving resolution, an angle on
the 2D mesh corresponds to a comoving length that varies with the
redshift.
For example, at low $z$, $\restwod \chi(z) \ll \resthreed$.
As a result, a 2D ray mesh samples forces from the 3D particle mesh at a
much higher resolution the latter doesn't offer.
So maintaining a high resolution 2D mesh at low redshift is both
computationally inefficiency.
Therefore, we employ an adaptive ray mesh, where we progressively
coarsen $\restwod$ towards low redshifts.
The choice of $\restwod$ depends on the 3D mesh resolution, pixel size,
and efficient FFT sizes (including padding).
Further details and convergence tests are discussed in
\cref{app:ray_mesh}.

\section{Validations}
\label{sec:validations}

\subsection{Lensing by a point mass}
\label{sec:pointmass}

\begin{figure}[ht]
\centering
\includegraphics[width=0.8\hsize,trim={0 11cm 0 0},clip]{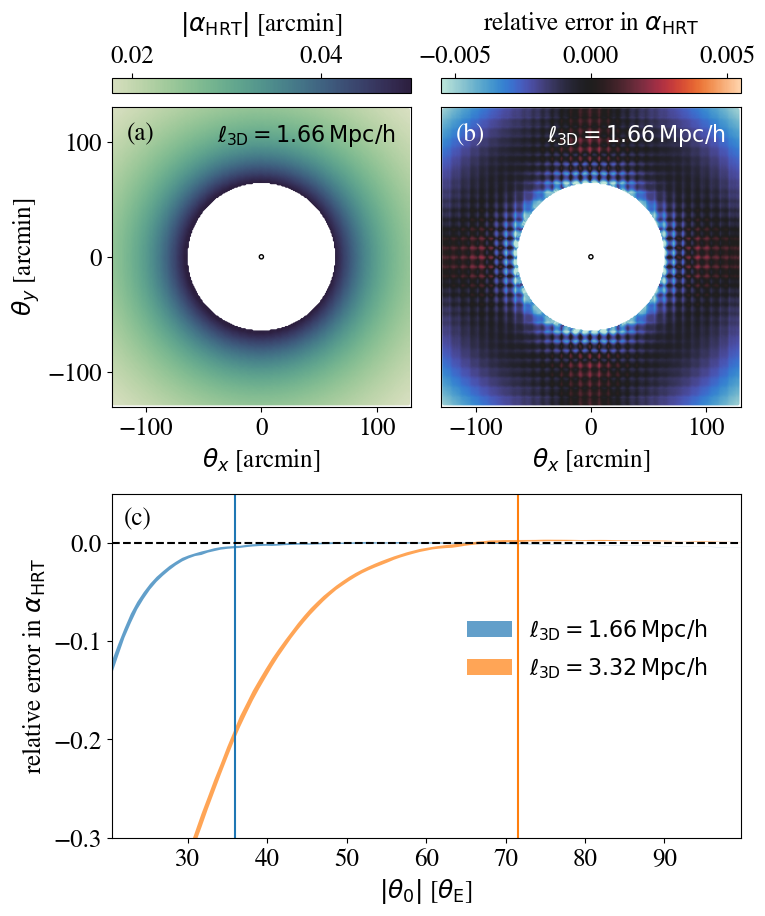}
\caption{We apply HRT to the problem of point mass lensing as described
in \cref{sec:pointmass}, using a $256 \times 256 \times 512$ box with a
mesh resolution $\resthreed = 1.66\,\Mpch$.
We show $\alphahrt$ in panel (a).
For a given $\resthreed$, the point mass angular resolution limit is
$\thetathresh = 4 \, \resthreed / \chi_\mathrm{l}$ (see
\cref{fig:pointmass_resolution_mass}).
We mask light rays within this threshold where the potential field of
the point mass lens is not well resolved.
The black circle indicates the Einstein radius $\thetae$.
Panel (b) illustrates the relative error between $\alphahrt$ and
$\alphatheory$.
HRT achieves accuracy within $<0.5\%$ for both configurations above the
resolution limit.}
\label{fig:pointmass_resolution}
\end{figure}

We test our ray tracing algorithm on the classic problem of
gravitational lensing by a point mass in a flat, static universe.
The observer is positioned at $\chi = 0$, with a lens mass $M = 1.3
\times 10^{15} \Msun$ (with an Einstein radius of $\thetae=1.8'$,
representative of a massive cluster) at $\chi_\mathrm{l} = 350~\Mpch$,
and the source plane at $\chi_\mathrm{s} = 850~\Mpch$ (or $z_\mathrm{s}
\approx 0.3$).
The theoretical prediction for the deflection angle $\alphatheory \equiv
\bftheta_\mathrm{theory} - \bftheta_0$ is provided by
\cite{epstein_post-post-newtonian_1980}:
\begin{equation}
\label{eqn:pointmass_analytic}
\alphatheory
= \frac{\chi_\mathrm{s} - \chi_\mathrm{l}}{\chi_\mathrm{s}}
\Bigl[
\hat\alpha
+ \frac{15\pi}{4} \hat\alpha^2
+ \mathcal{O}\left(\hat\alpha^3\right)
\Bigr]
\frac{\bftheta_0}{\left|\bftheta_0\right|} \,,
\end{equation}
where $\hat\alpha \equiv 4GM / (b c^2)$ and $b = \chi_\mathrm{l}
\left|\bftheta_0\right|$ is the impact parameter.
We solve the same problem using HRT.
We define a 3D mesh of size $(256, 256, 512)$ with a resolution of
$1.66~\Mpch$, and an image plane spanning $256' \times 256'$ with a
pixel resolution of $1'$.
We initialize rays on a uniform grid ($\bftheta_0$) at $z=0$ and trace
them to $z_\mathrm{s}$ in $45$ time steps.
Panel (a) of \cref{fig:pointmass_resolution} shows the deflection angle
$\alphahrt$ obtained via HRT.

\begin{figure}[ht]
\centering
\includegraphics[width=0.85\hsize]{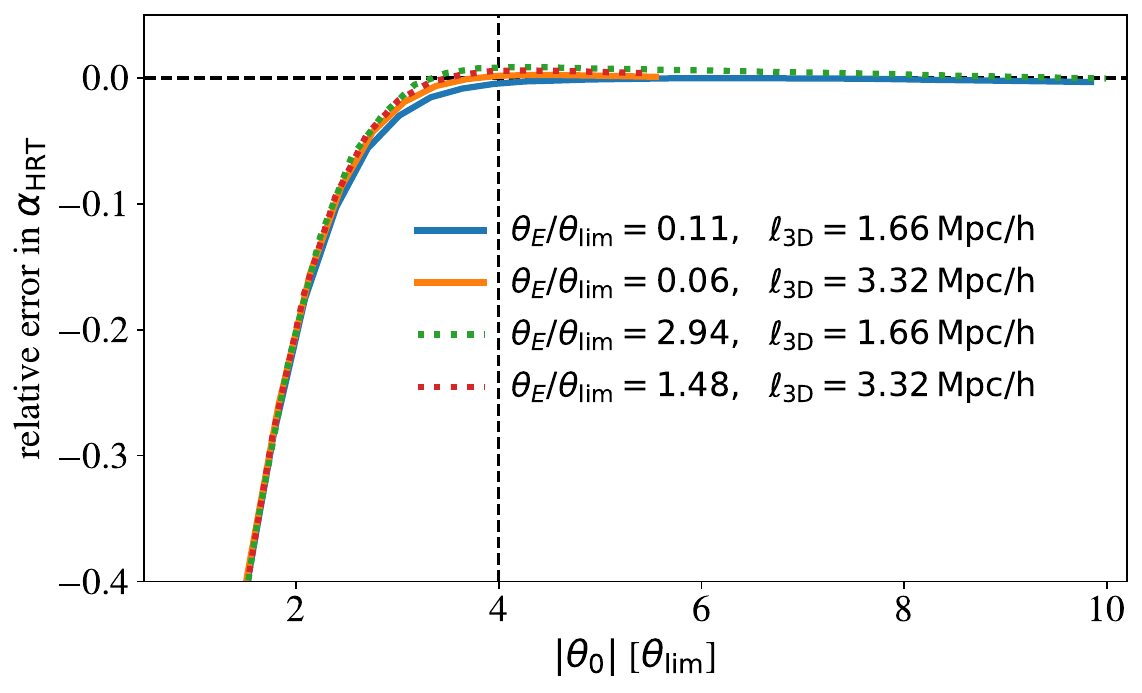}
\caption{Relative error of $\alphahrt$ versus $|\theta_0|$ (in unit of
$\thetathresh$) for two different 3D mesh resolutions and two different
lens masses.
The low and high mass cases are shown in solid and dashed lines,
respectively, with their masses represented in terms of their Einstein
radiuses $\thetae$.
We observe a universal relationship between the accuracy and
$|\bftheta_0|/\thetathresh$.
HRT maintains sub-percent accuracy beyond $\thetathresh$, demonstrating
that mesh resolution is the primary determinant of accuracy.}
\label{fig:pointmass_resolution_mass}
\end{figure}

The HRT result aligns with theoretical predictions with very high
accuracy.
We plot the relative error on the image plane, defined by $\epsilon =
\lvert\alphahrt - \alphatheory\rvert / \lvert \alphatheory\rvert$, in
panel (b) of \cref{fig:pointmass_resolution}.
HRT consistently achieves accuracy within $0.5\%$ across the image plane
where the 3D mesh resolution is adequate.
Specifically, we mask pixels within $\thetathresh = 4 \, \resthreed /
\chi_\mathrm{l}$, where the potential field generated by the point
particle cannot be clearly resolved due to the finite resolution of the
3D mesh.
A lower resolution mesh dampens the potential field at the mesh scale,
thereby suppressing $\alphahrt$ near the lens mass.
\Cref{fig:pointmass_resolution_mass} illustrates $\epsilon$ for the $M =
1.3 \times 10^{15} \Msun$ case above as a function of $\thetathresh$ for
two configurations with different resolutions: $\resthreed = 1.66~\Mpch$
and $3.32~\Mpch$.
These two configurations test HRT in the weak field limit, where the
Einstein radius $\thetae \ll \thetathresh$.
As expected, for pixels falling within $\thetathresh$, $\alphahrt$ is
systematically lowered (\cref{fig:pointmass_resolution_mass}, solid blue
line).
Halving the 3D mesh resolution doubles $\thetathresh$, but $\alphahrt$
still demonstrates percent-level accuracy outside $\thetathresh$ and is
suppressed within it (\cref{fig:pointmass_resolution_mass}, solid orange
line).
We also test HRT in a stronger gravitational field by increasing $M$
such that $\thetae \approx \thetathresh$.
The result for the high and low resolution cases are shown in dashed
green and red lines in \cref{fig:pointmass_resolution_mass},
respectively.
In general, we find the definition of $\thetathresh$ serves as a robust
and universal threshold to characterize the accuracy of HRT regardless
of $\resthreed$ and $M$.
This also shows that HRT is accurate as long as the 3D mesh resolution
is sufficiently high.
The accuracy of HRT also weakly depends on the resolution and boundary
conditions of the ray mesh, as well as the number of time steps.
We characterize these effects in \cref{sec:point_mass_error}.

\subsection{Post-Born weak lensing in a cosmological volume}
\label{sec:cl}

\begin{figure}
\centering
\includegraphics[width=0.8\hsize]{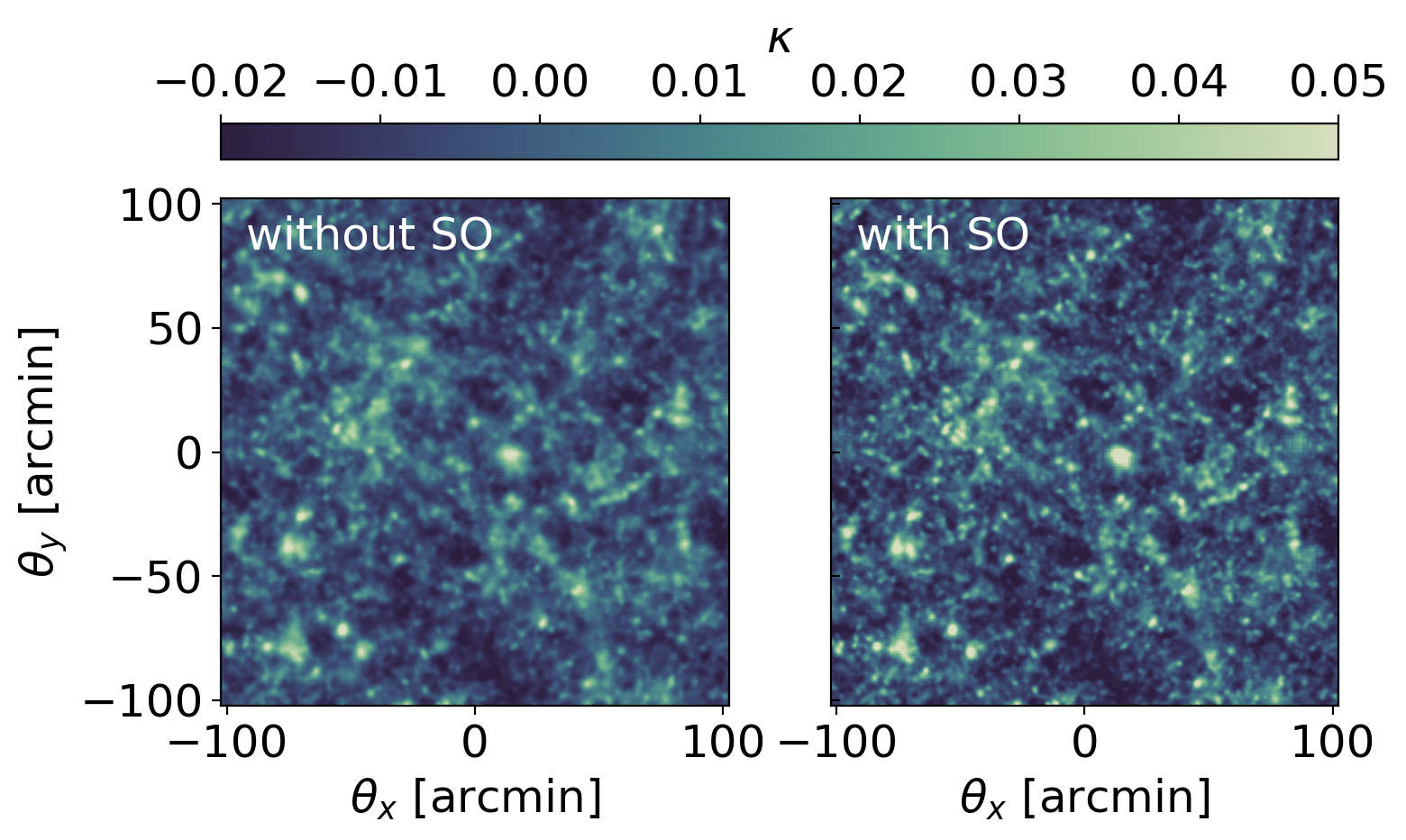}
\caption{Convergence ($\kappa$) maps generated using the HRT algorithm
on a cosmological simulation.
The left panel shows the $\kappa$ map derived from a density field
evolved using the standard PM gravity solver.
The right panel illustrates the $\kappa$ map derived from the same
initial cosmological conditions but evolved using the PM algorithm
enhanced with the spatial optimization (SO) gravity solver, that
sharpens the convergence peaks compared to the regular PM.}
\label{fig:kappa}
\end{figure}

\begin{figure}
    \centering
    \includegraphics[width=0.7\hsize]{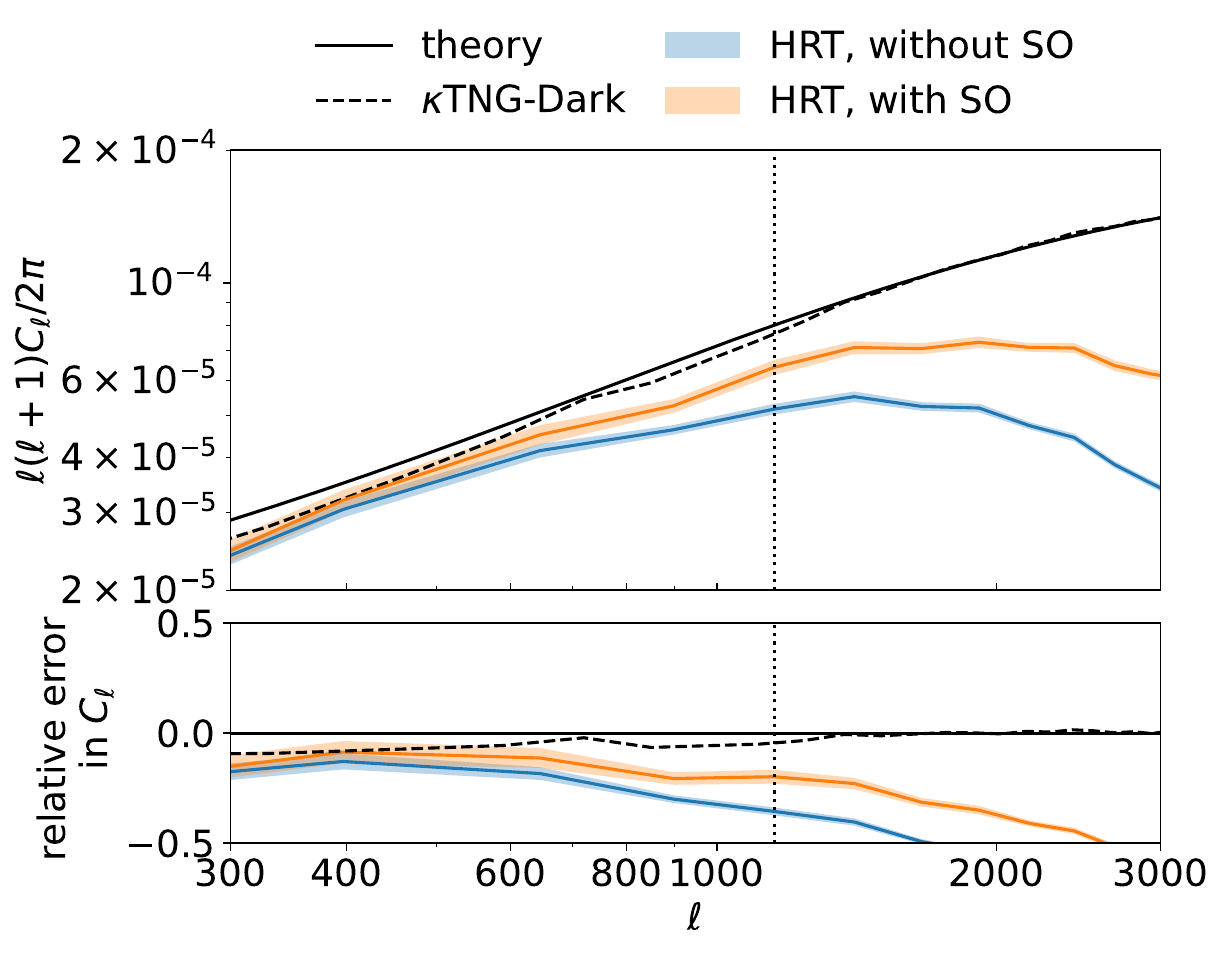}
    \caption{The top panel shows the distribution of the convergence
    power spectra obtained by ray tracing through cosmological
    simulations using HRT. The bottom panel shows the relative error
    compared to theory (black, solid) for the same cosmology and $n(z)$.
    The $\ktng$ results are shown in black, dashed lines for comparison.
    Simulations where gravity is solved using the standard PM method are
    depicted in blue, while those using the PM enhanced with the spatial
    optimization (SO) force-sharpening method are shown in orange. The HRT and the $\ktng$ $C_\ell$'s are similarly suppressed at low $\ell$'s because of the finite simulation box size and the lack of large-scale modes.
    The HRT results align with $\ktng$ at large scales but are
    suppressed at smaller scales since PM methods cannot resolve
    interactions at or below the mesh resolution.
    Enhancing PM with SO mitigates this issue and extends $\chrt$ to
    higher $\ell$'s.
    For the PM+SO result \response{(where the redshift distribution is a delta function at the same redshift, $z=1.034$, as the $\ktng$ lensing map)}, $\chrt$ agrees with $\cktng$ to within $10\%$ for $\ell<800$ and $15\%$ for $\ell<1200$, where $\ell=1200$ is marked by the vertical dotted line.}
    \label{fig:cl}
\end{figure}

We use the HRT algorithm to perform ray tracing over a cosmological
volume and study the statistics of weak lensing observables.
Throughout this paper, we adopt the \emph{Planck} 2015 cosmology
\cite{planck_collaboration_planck_2016}.
This cosmology also underpins the $\ktng$ simulations
\cite{osato_kappatng_2021} and the lensing forward model by Ref.~\cite{lanzieri_forecasting_2023}, with which we compare our results.
The $\ktng$ simulations are obtained by post-processing the higher
resolution dark matter-only TNG300-1-Dark simulations (hereafter
TNG-Dark) using the MLP ray tracing algorithm \cite{osato_kappatng_2021,
naiman_first_2018, springel_first_2018, marinacci_first_2018,
nelson_first_2018, pillepich_first_2018}.
Ref.~\cite{lanzieri_forecasting_2023} constructed weak lensing
convergence maps with PM simulation using the Born approximation.
The main purpose of this paper is to present the HRT algorithm itself;
we will focus on power spectrum recovery here and leave detailed
higher-order statistics analysis for a future work.

Our fiducial results are obtained using a $512 \times 512 \times 512$
simulation box with a particle spacing of $0.4~\Mpch$ and a mesh spacing of $0.4~\Mpch$.
We first evolve the particles from initial condition to $a=1/64$ using
2nd-order Lagrangian perturbation theory.
We then simulate the gravitational interaction using the PM algorithm
from $a=1/64$ to today in 64 time steps\footnote{This choice differs
from Ref.~\cite{lanzieri_forecasting_2023}'s, who seeds the initial
condition at $a=1/7$.}.
From there, we perform ray tracing back to $z_s = 1.034$ across $30$
time steps. The lens distirbution is the entire density field in the simulation, and the source distribution is a Dirac delta distribution at $z_s$.
The image plane spans $205' \times 205'$; it includes $1024\times1024$
pixels with a $0.2'$ pixel size.
The size and resolution of our simulation box are constrained by the
memory capacity of the GPU\footnote{For this test, we use a H100 NVL GPU on the Vera cluster at the
Pittsburgh Supercomputing Center.}.
Consequently, we do not yet have the hardware capability to conduct ray
tracing within a single, monolithic simulation box.
Instead, we tile our past light cone by replicating the snapshots $12$
times, each modified by a random translation and rotation along the
three spatial axes.
This tiling strategy, extensively studied in Ref.~\cite{petri_sample_2016},
can produce $10^4$ independent realizations of weak lensing power
spectra and peak statistics with even a single snapshot.
This process has been used in many weak lensing mocks including the
$\ktng$ simulations \cite{osato_kappatng_2021}.
An example of our ray-traced $\kappa$ map is shown in the left panel of
\cref{fig:kappa}.
We run $50$ independent simulations and present the distribution of
$\chrt$ along with its comparison to theory (black, solid) and $\cktng$ (black, dashed) in the top and bottom
panels of \cref{fig:cl} (in blue).
The plot shows that $\chrt$ aligns with $\cktng$ within $20\%$ for $\ell
< 800$ and within $30\%$ for $\ell < 1200$.
$\chrt$ is suppressed on the large scale because our simulation box is
small and can only include a limited number of large-scale modes.
However, the amount of deficit in power is consistent with the result
$\cktng$ and the results in obtained by
Ref.~\cite{lanzieri_forecasting_2023} who performed lensing simulation
using the same simulation volume.
We also observe $\chrt$ to be suppressed at high $\ell$'s.
This is because the PM gravity solver cannot accurately resolve
gravitational interactions at the mesh scale.
This result is comparable to the ``DLL without correction'' analysis in
Ref.~\cite{lanzieri_forecasting_2023}, who recovers lensing $C_\ell$ up
to $\ell \approx 300$ compared to $\cktng$.
We recover $C_\ell$ over a larger $\ell$ range because of the finer mesh
size used in the present work.

\begin{figure}
    \centering
    \includegraphics[width=0.7\hsize]{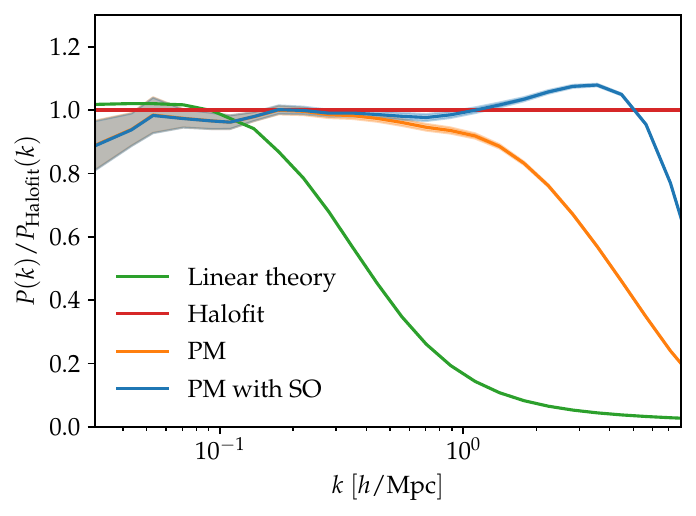}
    \caption{Matter power spectrum at $z=0$ with the regular PM solver
    (orange) and with PM + spatial optimization (SO, blue).
    They are compared to the prediction of the Halofit model in red and
    the linear theory in green.
    The shaded regions represent the standard error of the mean
    estimated with 50 independent simulations.
    The simulation has $512^3$ particles with a $0.4~\Mpch$ particle
    spacing.
    SO corrects for the small-scale interactions between dark matter
    particles that are ignored by the regular PM, boosting power at
    high-$k$.}
    \label{fig:matter_power_so}
\end{figure}

Next, we explore the potential of recovering the convergence power
spectrum at higher $\ell$ ranges by extending the regular PM gravity
solver to smaller scales with a spatial optimization (SO) algorithm
that sharpens the PM forces near the mesh scale \cite{ZhangLiEtAl}.
In Fourier space, the gravitational force is proportional to $(\bfk/k^2)
\delta(\bfk)$.
SO modifies the Fourier force kernel by a nonlinear function $g$:
\begin{equation}
\frac{\bfk}{k^2} \rightarrow \frac{\bfk}{k^2} g(\bfk; \bm{\vartheta}) \,,
\end{equation}
where $\bm{\vartheta}$ includes cosmological parameters and simulation
configurations.
We implement $g$ using symmetry-preserving neural networks, which are
trained to align the PM+SO simulations with the GADGET-4
\cite{springel_simulating_2021} simulations across a wide range of
cosmologies and simulation configurations.
As illustrated in \cref{fig:matter_power_so}, SO boosts the small-scale
$P(k)$ compared to the regular PM algorithm for $k > 1 h/\mathrm{Mpc}$,
though still slightly lower than the halofit
\cite{takahashi_revising_2012} predictions.
We run HRT simulations under the same settings as previously discussed
but integrate SO with the same initial conditions.
An example $\kappa$ map, shown in the right panel of \cref{fig:kappa},
indicates that SO indeed sharpens the small-scale fluctuations compared
to the regular PM simulations.
The distribution of power spectra with SO is displayed in \cref{fig:cl}
in orange. SO consistently improves the convergence power spectrum at all scales, but is especially helpful at high $\ell$'s. $\chrt$ agrees with $\cktng$ to within $10\%$ for $\ell<800$ and $15\%$ for $\ell<1200$.
In general, the maximum $\ell$ at which we can recover $C_\ell$ depends
on the 3D mesh resolution $\resthreed$ \response{and the shape of the lensing kernel.
Empirically, we find this scale to be approximately $\ell \approx
\frac{\ell_\mathrm{lim}}{16} \approx 1200$ for $z_s=1.034$ when SO is applied
(vertical dotted line in \cref{fig:cl}). We define $\ell_\mathrm{lim} =
\pi / \left(\frac{\chi(z_s)}{2\resthreed} \right)$, since $\frac{\chi(z_s)}{2\resthreed}$ is the angular scale of the 3D mesh at the radial comoving distance where the lensing kernel has the most sensitivity. More generally, when the source distribution is not a delta function, the denominator of $\ell_\mathrm{lim}$ should be similarly approximated by the distance to the peak of the lensing kernel.}

We anticipate that higher mesh resolution and stronger force-sharpening
effects will enable $\chrt$ to reach even smaller scales, which is
potentially achievable when larger GPU memory becomes accessible in the future.
We also conduct convergence tests with varying HRT hyper-parameters in
\cref{app:cl}.

\section{Conclusions}
\label{sec:conclusion}

We have presented a post-Born, three-dimensional, on-the-fly ray tracing
algorithm based on the Hamiltonian dynamics of light rays.
This method, termed Hamiltonian ray tracing (HRT), includes the lens-lens coupling and does not assume the Born approximation.
Additionally, HRT deflects photons based on the gravitational potential
generated by the entire cosmological volume, not just that within a
single lens plane.
HRT also performs ray tracing on-the-fly. As a result, it integrates well with
any gravity solver and computes light ray trajectories and lensing
observable maps with minimal computational overhead compared to running
the gravity solver alone.
We implemented HRT using the \pmwd\ library on a GPU platform,
demonstrating its accuracy and limitations both for point mass lensing
and in generating convergence maps and power spectra for dark matter
simulations.
For point-mass lensing, HRT yields deflection angles accurate to
sub-percent levels above the resolution limit.
When applied to cosmological simulations using the standard PM gravity
solver with a $0.4~\Mpch$ particle spacing, HRT generates lensing
convergence maps whose power spectrum aligns with the fiducial $\ktng$
results to within $20\%$ of $\cktng$ for $\ell<800$ and $30\%$ for $\ell<1200$.
Extending the PM with small-scale force-sharpening method (SO) enables
recovery of $C_\ell$ to within $10\%$ of $\cktng$ for $\ell<800$ and $15\%$ for $\ell<1200$. 
\response{Future works could extend the tests to other scenarios, such as lensing by compensated black holes in an evolving cosmological setting \cite{Hu_2022} or different weak lensing statistics paired with various redshift distributions.}

While HRT should work with any simulation of structure formation, its
implementation is particularly easy in frameworks with automatic
differentiation capability, such as \pmwd\ which is based on \jax.
The forward mode in automatic differentiation helps to co-evolve the
lensing observables, such as cosmic shear, with the light ray
deflections, since the Jacobian of the latter involves the former.
On the other hand, we would also need the reverse-mode differentiation
to compute the likelihood gradient, for example, in FLI applications
that involve Hamiltonian Monte Carlo.
Automatic reverse-mode differentiation through the whole simulation can
be extremely memory consuming, and the adjoint method \cite{LiEtAl2024}
has been introduced to obtain memory-efficient gradients.
The same method can be applied to HRT, which we leave for future
development.

The accuracy of HRT is primarily limited by the mesh resolution and the
precision of the gravity solver at small comoving scales.
If we assume that the smallest scale at which the gravity solver is
accurate is $k_\mathrm{max} \propto \resthreed^{-1}$, and that the
smallest angular scale at which HRT is accurate is $\ell_\mathrm{max}
\propto \resthreed^{-1}$, then $\ell_\mathrm{max} \propto
k_\mathrm{max}$.
For the current generation of weak lensing surveys like HSC,
$\ell_\mathrm{max} \approx 1800$ \cite{dalal_hyper_2023}, about 2 times
higher than the $\ell$ limit we achieve here.
To bridge the gap between simulation and observational data, future work
could extend the SO framework to push $k_\mathrm{max}$ above 5 $h/$Mpc where baryonic effects also becomes important. 
Alternatively, increasing the resolution of the cosmological simulation
by two or three-folds, which could be achieved by parallelizing the PM
code and the HRT algorithm across multiple GPU devices or nodes, may
also prove equally effective.

The HRT algorithm empowers cosmological analysis in several ways.
HRT can quickly generate cosmology-dependent ray tracing shear maps,
making it suitable for training machine learning models for
simulation-based inference.
It can help establishing the connection between cosmology and
higher-order statistics in a data-driven manner and aid in estimating
cosmology-dependent covariance matrices for summary statistics analyses.
As a differentiable ray tracing algorithm, HRT also enables field-level
inference that accounts for post-Born effects.
While this work focuses on the algorithm and its implementation, future
studies could extend this discussion with a detailed analysis of
higher-order statistics in the convergence maps simulated by HRT.
Additionally, optimizing the differentiation of HRT through the adjoint
method could make it more memory-friendly in field-level inference
applications.

\section*{Code availability}
\pmwd\ is open-source on GitHub
\href{https://github.com/eelregit/pmwd}{\faGithub}.
The ray tracing feature will be made available in that repository after
code cleaning, including the source files and scripts of this paper
\href{https://github.com/eelregit/pmwd/tree/master/docs/papers/hrt}{\faFile}.

\section*{Acknowledgement}
AZ thanks Yuuki Omori for discussion on post-born validations on weak
lensing. 
YL thanks Sukhdeep Singh for helpful discussion. 
AZ, XL, and RM were partially supported by a grant from the Simons Foundation (Simons Investigator in Astrophysics, Award ID 620789).
YL and YZ are supported by The Major Key Project of PCL.
YZ is further supported by the China Postdoctoral Science Foundation
under award number 2023M731831. 
GF acknowledges the support of the European Research Council under the Marie Sk\l{}odowska Curie actions through the Individual Global Fellowship No.~892401 PiCOGAMBAS and of the Simons Foundation for the initial stages of this work.
This work is supported by the Bridges-2 supercomputer at the Pittsburgh
Supercomputing Center under the NSF ACCESS Explore allocation PHY230147.
We thank the Columbia Lensing group (http://columbialensing.org) for
making their simulations available. 
The creation of these simulations is supported through grants NSF
AST-1210877, NSF AST-140041, and NASA ATP-80NSSC18K1093. 

\bibliographystyle{JHEP}
\bibliography{references, references_yin, references_giulio, references_others}

\providecommand{\href}[2]{#2}\begingroup\raggedright\begin{thebibliography}{10}

\bibitem{KiesslingEtAl2011}
A.~Kiessling, A.F.~Heavens, A.N.~Taylor and B.~Joachimi, \emph{Sunglass: A new
  weak-lensing simulation pipeline},
  \href{https://doi.org/10.1111/j.1365-2966.2011.18540.x}{\emph{Monthly Notices
  of the Royal Astronomical Society} {\bfseries 414} (2011) 2235}.

\bibitem{FosalbaEtAl2015}
P.~Fosalba, E.~Gazta{\~n}aga, F.J.~Castander and M.~Crocce, \emph{The {{MICE
  Grand Challenge}} light-cone simulation -- {{III}}. {{Galaxy}} lensing mocks
  from all-sky lensing maps},
  \href{https://doi.org/10.1093/mnras/stu2464}{\emph{Monthly Notices of the
  Royal Astronomical Society} {\bfseries 447} (2015) 1319}.

\bibitem{SgierEtAl2019}
R.J.~Sgier, A.~R{\'e}fr{\'e}gier, A.~Amara and A.~Nicola, \emph{Fast generation
  of covariance matrices for weak lensing},
  \href{https://doi.org/10.1088/1475-7516/2019/01/044}{\emph{Journal of
  Cosmology and Astroparticle Physics} {\bfseries 2019} (2019) 044}.

\bibitem{TessoreEtAl2023}
N.~Tessore, A.~Loureiro, B.~Joachimi, M.~{von Wietersheim-Kramsta} and
  N.~Jeffrey, \emph{{{GLASS}}: Generator for large scale structure},
  \href{https://doi.org/10.21105/astro.2302.01942}{\emph{The Open Journal of
  Astrophysics} {\bfseries 6} (2023) 10.21105/astro.2302.01942}
  [\href{https://arxiv.org/abs/2302.01942}{{\ttfamily 2302.01942}}].

\bibitem{li_constraining_2019}
Z.~Li, J.~Liu, J.M.Z.~Matilla and W.R.~Coulton, \emph{Constraining neutrino
  mass with tomographic weak lensing peak counts},
  \href{https://doi.org/10.1103/PhysRevD.99.063527}{\emph{Physical Review D}
  {\bfseries 99} (2019) 063527}.

\bibitem{liu_cosmology_2015}
J.~Liu, A.~Petri, Z.~Haiman, L.~Hui, J.M.~Kratochvil and M.~May,
  \emph{Cosmology {Constraints} from the {Weak} {Lensing} {Peak} {Counts} and
  the {Power} {Spectrum} in {CFHTLenS}},
  \href{https://doi.org/10.1103/PhysRevD.91.063507}{\emph{Physical Review D}
  {\bfseries 91} (2015) 063507}.

\bibitem{cranmer_frontier_2020}
K.~Cranmer, J.~Brehmer and G.~Louppe, \emph{The frontier of simulation-based
  inference}, \href{https://doi.org/10.1073/pnas.1912789117}{\emph{Proceedings
  of the National Academy of Sciences} {\bfseries 117} (2020) 30055}.

\bibitem{LiEtAl2024}
Y.~Li, C.~Modi, D.~Jamieson, Y.~Zhang, L.~Lu, Y.~Feng et~al.,
  \emph{Differentiable cosmological simulation with the adjoint method},
  \href{https://doi.org/10.3847/1538-4365/ad0ce7}{\emph{The Astrophysical
  Journal Supplement Series} {\bfseries 270} (2024) 36}
  [\href{https://arxiv.org/abs/2211.09815}{{\ttfamily 2211.09815}}].

\bibitem{pmwd}
Y.~Li, L.~Lu, C.~Modi, D.~Jamieson, Y.~Zhang, Y.~Feng et~al., \emph{\pmwd: A
  differentiable cosmological particle-mesh {{N-body}} library}, .

\bibitem{zhou_field-level_2023}
A.J.~Zhou and S.~Dodelson, \emph{Field-level multiprobe analysis of the {CMB},
  integrated {Sachs}-{Wolfe} effect, and the galaxy density maps},
  \href{https://doi.org/10.1103/PhysRevD.108.083506}{\emph{Physical Review D}
  {\bfseries 108} (2023) 083506}.

\bibitem{zhou_accurate_2023}
A.J.~Zhou, X.~Li, S.~Dodelson and R.~Mandelbaum, \emph{Accurate field-level
  weak lensing inference for precision cosmology},
  \href{https://doi.org/10.1103/PhysRevD.110.023539}{\emph{Physical Review D}
  {\bfseries 110} (2024) 023539}.

\bibitem{BORG}
J.~Jasche and B.D.~Wandelt, \emph{Bayesian physical reconstruction of initial
  conditions from large-scale structure surveys},
  \href{https://doi.org/10.1093/mnras/stt449}{\emph{Monthly Notices of the
  Royal Astronomical Society} {\bfseries 432} (2013) 894}
  [\href{https://arxiv.org/abs/1203.3639}{{\ttfamily 1203.3639}}].

\bibitem{ELUCID}
H.~Wang, H.J.~Mo, X.~Yang, Y.P.~Jing and W.P.~Lin, \emph{{{ELUCID}} - exploring
  the local universe with the reconstructed initial density field. {{I}}.
  hamiltonian markov chain monte carlo method with particle mesh dynamics},
  \href{https://doi.org/10.1088/0004-637X/794/1/94}{\emph{The Astrophysical
  Journal} {\bfseries 794} (2014) 94}
  [\href{https://arxiv.org/abs/1407.3451}{{\ttfamily 1407.3451}}].

\bibitem{SeljakEtAl2017}
U.~Seljak, G.~Aslanyan, Y.~Feng and C.~Modi, \emph{Towards optimal extraction
  of cosmological information from nonlinear data},
  \href{https://doi.org/10.1088/1475-7516/2017/12/009}{\emph{Journal of
  Cosmology and Astroparticle Physics} {\bfseries 2017} (2017) 009}
  [\href{https://arxiv.org/abs/1706.06645}{{\ttfamily 1706.06645}}].

\bibitem{BORG-PM}
J.~Jasche and G.~Lavaux, \emph{Physical {{Bayesian}} modelling of the
  non-linear matter distribution: New insights into the nearby universe},
  \href{https://doi.org/10.1051/0004-6361/201833710}{\emph{Astronomy \&
  Astrophysics} {\bfseries 625} (2019) A64}
  [\href{https://arxiv.org/abs/1806.11117}{{\ttfamily 1806.11117}}].

\bibitem{SchmidtEtAl2019}
F.~Schmidt, F.~Elsner, J.~Jasche, N.M.~Nguyen and G.~Lavaux, \emph{A rigorous
  {EFT}-based forward model for large-scale structure}, {\emph{Journal of
  Cosmology and Astroparticle Physics} {\bfseries 2019} (2019) 042}.

\bibitem{alsing_cosmological_2017}
J.~Alsing, A.F.~Heavens and A.H.~Jaffe, \emph{Cosmological parameters, shear
  maps and power spectra from {CFHTLenS} using {Bayesian} hierarchical
  inference}, \href{https://doi.org/10.1093/mnras/stw3161}{\emph{Monthly
  Notices of the Royal Astronomical Society} {\bfseries 466} (2017) 3272}.

\bibitem{anderes_bayesian_2015}
E.~Anderes, B.~Wandelt and G.~Lavaux, \emph{Bayesian inference of {CMB}
  gravitational lensing},
  \href{https://doi.org/10.1088/0004-637X/808/2/152}{\emph{The Astrophysical
  Journal} {\bfseries 808} (2015) 152}.

\bibitem{KitauraEtAl2021}
F.-S.~Kitaura, M.~Ata, S.A.~Rodr{\'\i}guez-Torres,
  M.~Hern{\'a}ndez-S{\'a}nchez, A.~Balaguera-Antol{\'\i}nez and G.~Yepes,
  \emph{Cosmic birth: efficient bayesian inference of the evolving cosmic web
  from galaxy surveys}, {\emph{Monthly Notices of the Royal Astronomical
  Society} {\bfseries 502} (2021) 3456}.

\bibitem{AtaEtAl2021}
M.~Ata, F.-S.~Kitaura, K.-G.~Lee, B.C.~Lemaux, D.~Kashino, O.~Cucciati et~al.,
  \emph{Birth of the cosmos field: primordial and evolved density
  reconstructions during cosmic high noon}, {\emph{Monthly Notices of the Royal
  Astronomical Society} {\bfseries 500} (2021) 3194}.

\bibitem{fiedorowicz_karmma_2022}
P.~Fiedorowicz, E.~Rozo, S.S.~Boruah, C.~Chang and M.~Gatti, \emph{{KaRMMa} --
  {Kappa} {Reconstruction} for {Mass} {Mapping}},
  \href{https://doi.org/10.1093/mnras/stac468}{\emph{Monthly Notices of the
  Royal Astronomical Society} {\bfseries 512} (2022) 73}.

\bibitem{fiedorowicz_karmma_2022-1}
P.~Fiedorowicz, E.~Rozo and S.S.~Boruah, \emph{{KaRMMa} 2.0 -- {Kappa}
  {Reconstruction} for {Mass} {Mapping}},  Oct., 2022.

\bibitem{millea_optimal_2021}
M.~Millea, C.M.~Daley, T.-L.~Chou, E.~Anderes, P.A.R.~Ade, A.J.~Anderson
  et~al., \emph{Optimal {CMB} {Lensing} {Reconstruction} and {Parameter}
  {Estimation} with {SPTpol} {Data}},
  \href{https://doi.org/10.3847/1538-4357/ac02bb}{\emph{The Astrophysical
  Journal} {\bfseries 922} (2021) 259}.

\bibitem{lanzieri_forecasting_2023}
D.~Lanzieri, F.~Lanusse, C.~Modi, B.~Horowitz, J.~Harnois-Déraps, J.-L.~Starck
  et~al., \emph{Forecasting the power of {Higher} {Order} {Weak} {Lensing}
  {Statistics} with automatically differentiable simulations},  May, 2023.

\bibitem{porqueres_field-level_2023}
N.~Porqueres, A.~Heavens, D.~Mortlock, G.~Lavaux and T.L.~Makinen,
  \emph{Field-level inference of cosmic shear with intrinsic alignments and
  baryons},  Apr., 2023.

\bibitem{NguyenEtAl2024}
N.-M.~Nguyen, F.~Schmidt, B.~Tucci, M.~Reinecke and A.~Kosti{\'c}, \emph{How
  much information can be extracted from galaxy clustering at the field
  level?}, {\emph{arXiv preprint arXiv:2403.03220} (2024) }.

\bibitem{dodelson_second_2005}
S.~Dodelson, E.W.~Kolb, S.~Matarrese, A.~Riotto and P.~Zhang, \emph{Second
  {Order} {Geodesic} {Corrections} to {Cosmic} {Shear}},
  \href{https://doi.org/10.1103/PhysRevD.72.103004}{\emph{Physical Review D}
  {\bfseries 72} (2005) 103004}.

\bibitem{cooray_second_2002}
A.~Cooray and W.~Hu, \emph{Second {Order} {Corrections} to {Weak} {Lensing} by
  {Large}-{Scale} {Structure}}, \href{https://doi.org/10.1086/340892}{\emph{The
  Astrophysical Journal} {\bfseries 574} (2002) 19}.

\bibitem{krause_hirata_2010}
E.~{Krause} and C.M.~{Hirata}, \emph{{Weak lensing power spectra for precision
  cosmology. Multiple-deflection, reduced shear, and lensing bias
  corrections}},
  \href{https://doi.org/10.1051/0004-6361/200913524}{\emph{Astronomy \&
  Astrophysics} {\bfseries 523} (2010) A28}
  [\href{https://arxiv.org/abs/0910.3786}{{\ttfamily 0910.3786}}].

\bibitem{pratten_impact_2016}
G.~Pratten and A.~Lewis, \emph{Impact of post-{Born} lensing on the {CMB}},
  \href{https://doi.org/10.1088/1475-7516/2016/08/047}{\emph{Journal of
  Cosmology and Astroparticle Physics} {\bfseries 2016} (2016) 047}.

\bibitem{fabbian_et_al_2018}
G.~{Fabbian}, M.~{Calabrese} and C.~{Carbone}, \emph{{CMB weak-lensing beyond
  the Born approximation: a numerical approach}},
  \href{https://doi.org/10.1088/1475-7516/2018/02/050}{\emph{jcap} {\bfseries
  2018} (2018) 050} [\href{https://arxiv.org/abs/1702.03317}{{\ttfamily
  1702.03317}}].

\bibitem{petri_validity_2017}
A.~Petri, Z.~Haiman and M.~May, \emph{Validity of the {Born} approximation for
  beyond {Gaussian} weak lensing observables},
  \href{https://doi.org/10.1103/PhysRevD.95.123503}{\emph{Physical Review D}
  {\bfseries 95} (2017) 123503}.

\bibitem{fabbian_et_al_2019}
G.~{Fabbian}, A.~{Lewis} and D.~{Beck}, \emph{{CMB lensing reconstruction
  biases in cross-correlation with large-scale structure probes}},
  \href{https://doi.org/10.1088/1475-7516/2019/10/057}{\emph{\jcap} {\bfseries
  2019} (2019) 057} [\href{https://arxiv.org/abs/1906.08760}{{\ttfamily
  1906.08760}}].

\bibitem{chang_delensing_2014}
C.~Chang and B.~Jain, \emph{Delensing {Galaxy} {Surveys}},
  \href{https://doi.org/10.1093/mnras/stu1104}{\emph{Monthly Notices of the
  Royal Astronomical Society} {\bfseries 443} (2014) 102}.

\bibitem{bohm_madlens_2020}
V.~Böhm, Y.~Feng, M.E.~Lee and B.~Dai, \emph{{MADLens}, a python package for
  fast and differentiable non-{Gaussian} lensing simulations},  Dec., 2020.

\bibitem{dodelson_reduced_2006}
S.~Dodelson, C.~Shapiro and M.~White, \emph{Reduced {Shear} {Power}
  {Spectrum}}, \href{https://doi.org/10.1103/PhysRevD.73.023009}{\emph{Physical
  Review D} {\bfseries 73} (2006) 023009}.

\bibitem{jain_raytracing_2000}
B.~Jain, U.~Seljak and S.~White, \emph{Ray‐tracing {Simulations} of {Weak}
  {Lensing} by {Large}‐{Scale} {Structure}},
  \href{https://doi.org/10.1086/308384}{\emph{The Astrophysical Journal}
  {\bfseries 530} (2000) 547}.

\bibitem{hirata_reconstruction_2003}
C.M.~Hirata and U.~Seljak, \emph{Reconstruction of lensing from the cosmic
  microwave background polarization},
  \href{https://doi.org/10.1103/PhysRevD.68.083002}{\emph{Physical Review D}
  {\bfseries 68} (2003) 083002}.

\bibitem{hilbert_accuracy_2020}
S.~Hilbert, A.~Barreira, G.~Fabbian, P.~Fosalba, C.~Giocoli, S.~Bose et~al.,
  \emph{The {Accuracy} of {Weak} {Lensing} {Simulations}},
  \href{https://doi.org/10.1093/mnras/staa281}{\emph{Monthly Notices of the
  Royal Astronomical Society} {\bfseries 493} (2020) 305}.

\bibitem{beck_et_al_2018}
D.~{Beck}, G.~{Fabbian} and J.~{Errard}, \emph{{Lensing reconstruction in
  post-Born cosmic microwave background weak lensing}},
  \href{https://doi.org/10.1103/PhysRevD.98.043512}{\emph{prd} {\bfseries 98}
  (2018) 043512} [\href{https://arxiv.org/abs/1806.01216}{{\ttfamily
  1806.01216}}].

\bibitem{bohm_et_al_2018}
V.~{B{\"o}hm}, B.D.~{Sherwin}, J.~{Liu}, J.C.~{Hill}, M.~{Schmittfull} and
  T.~{Namikawa}, \emph{{Effect of non-Gaussian lensing deflections on CMB
  lensing measurements}},
  \href{https://doi.org/10.1103/PhysRevD.98.123510}{\emph{prd} {\bfseries 98}
  (2018) 123510} [\href{https://arxiv.org/abs/1806.01157}{{\ttfamily
  1806.01157}}].

\bibitem{bohm_et_al_2020}
V.~{B{\"o}hm}, C.~{Modi} and E.~{Castorina}, \emph{{Lensing corrections on
  galaxy-lensing cross correlations and galaxy-galaxy auto correlations}},
  \href{https://doi.org/10.1088/1475-7516/2020/03/045}{\emph{jcap} {\bfseries
  2020} (2020) 045} [\href{https://arxiv.org/abs/1910.06722}{{\ttfamily
  1910.06722}}].

\bibitem{vale_simulating_2003}
C.~Vale and M.~White, \emph{Simulating {Weak} {Lensing} by {Large}‐{Scale}
  {Structure}}, \href{https://doi.org/10.1086/375867}{\emph{The Astrophysical
  Journal} {\bfseries 592} (2003) 699}.

\bibitem{hilbert_strong_2007}
S.~Hilbert, S.D.M.~White, J.~Hartlap and P.~Schneider, \emph{Strong lensing
  optical depths in a {CDM} universe},
  \href{https://doi.org/10.1111/j.1365-2966.2007.12391.x}{\emph{Monthly Notices
  of the Royal Astronomical Society} {\bfseries 382} (2007) 121}.

\bibitem{hilbert_ray-tracing_2009}
S.~Hilbert, J.~Hartlap, S.D.M.~White and P.~Schneider, \emph{Ray-tracing
  through the {Millennium} {Simulation}: {Born} corrections and lens-lens
  coupling in cosmic shear and galaxy-galaxy lensing},
  \href{https://doi.org/10.1051/0004-6361/200811054}{\emph{Astronomy \&
  Astrophysics} {\bfseries 499} (2009) 31}.

\bibitem{sato_simulations_2009}
M.~Sato, T.~Hamana, R.~Takahashi, M.~Takada, N.~Yoshida, T.~Matsubara et~al.,
  \emph{Simulations of {Wide}-{Field} {Weak} {Lensing} {Surveys} {I}: {Basic}
  {Statistics} and {Non}-{Gaussian} {Effects}},
  \href{https://doi.org/10.1088/0004-637X/701/2/945}{\emph{The Astrophysical
  Journal} {\bfseries 701} (2009) 945}.

\bibitem{Becker2013}
M.R.~Becker, \emph{{{CALCLENS}}: Weak lensing simulations for large-area sky
  surveys and second-order effects in cosmic shear power spectra},
  \href{https://doi.org/10.1093/mnras/stt1352}{\emph{Monthly Notices of the
  Royal Astronomical Society} {\bfseries 435} (2013) 115}
  [\href{https://arxiv.org/abs/1210.3069}{{\ttfamily 1210.3069}}].

\bibitem{petri_mocking_2016}
A.~Petri, \emph{Mocking the {Weak} {Lensing} universe: the {LensTools} python
  computing package},
  \href{https://doi.org/10.1016/j.ascom.2016.06.001}{\emph{Astronomy and
  Computing} {\bfseries 17} (2016) 73}.

\bibitem{takahashi_full-sky_2017}
R.~Takahashi, T.~Hamana, M.~Shirasaki, T.~Namikawa, T.~Nishimichi, K.~Osato
  et~al., \emph{Full-sky {Gravitational} {Lensing} {Simulation} for
  {Large}-area {Galaxy} {Surveys} and {Cosmic} {Microwave} {Background}
  {Experiments}}, \href{https://doi.org/10.3847/1538-4357/aa943d}{\emph{The
  Astrophysical Journal} {\bfseries 850} (2017) 24}.

\bibitem{osato_kappatng_2021}
K.~Osato, J.~Liu and Z.~Haiman, \emph{\${\textbackslash}kappa\${TNG}: {Effect}
  of {Baryonic} {Processes} on {Weak} {Lensing} with {IllustrisTNG}
  {Simulations}}, \href{https://doi.org/10.1093/mnras/stab395}{\emph{Monthly
  Notices of the Royal Astronomical Society} {\bfseries 502} (2021) 5593}.

\bibitem{petri_et_al_2017}
A.~{Petri}, Z.~{Haiman} and M.~{May}, \emph{{Validity of the Born approximation
  for beyond Gaussian weak lensing observables}},
  \href{https://doi.org/10.1103/PhysRevD.95.123503}{\emph{prd} {\bfseries 95}
  (2017) 123503} [\href{https://arxiv.org/abs/1612.00852}{{\ttfamily
  1612.00852}}].

\bibitem{WeiEtAl2018}
C.~Wei, G.~Li, X.~Kang, Y.~Luo, Q.~Xia, P.~Wang et~al., \emph{Full-sky
  {{Ray-tracing Simulation}} of {{Weak Lensing Using ELUCID Simulations}}:
  {{Exploring Galaxy Intrinsic Alignment}} and {{Cosmic Shear Correlations}}},
  \href{https://doi.org/10.3847/1538-4357/aaa40d}{\emph{The Astrophysical
  Journal} {\bfseries 853} (2018) 25}.

\bibitem{XuJing2021}
K.~Xu and Y.~Jing, \emph{An accurate {{P3M}} algorithm for gravitational
  lensing studies in simulations},
  \href{https://doi.org/10.3847/1538-4357/ac0249}{\emph{The Astrophysical
  Journal} {\bfseries 915} (2021) 75}
  [\href{https://arxiv.org/abs/2102.08629}{{\ttfamily 2102.08629}}].

\bibitem{Jimenez-VicenteMediavilla2022}
J.~{Jim{\'e}nez-Vicente} and E.~Mediavilla, \emph{Fast multipole method for
  gravitational lensing: Application to high-magnification quasar
  microlensing}, \href{https://doi.org/10.3847/1538-4357/ac9e59}{\emph{The
  Astrophysical Journal} {\bfseries 941} (2022) 80}
  [\href{https://arxiv.org/abs/2211.00354}{{\ttfamily 2211.00354}}].

\bibitem{SuoEtAl2023}
X.~Suo, X.~Kang, C.~Wei and G.~Li, \emph{The spherical fast multipole method
  ({{sFMM}}) for gravitational lensing simulation},
  \href{https://doi.org/10.3847/1538-4357/acc107}{\emph{The Astrophysical
  Journal} {\bfseries 948} (2023) 56}
  [\href{https://arxiv.org/abs/2210.07021}{{\ttfamily 2210.07021}}].

\bibitem{CouchmanEtAl1999}
H.M.P.~Couchman, A.J.~Barber and P.A.~Thomas, \emph{Measuring the
  three-dimensional shear from simulation data, with applications to weak
  gravitational lensing},
  \href{https://doi.org/10.1046/j.1365-8711.1999.02714.x}{\emph{Monthly Notices
  of the Royal Astronomical Society} {\bfseries 308} (1999) 180}.

\bibitem{LiEtAl2011}
B.~Li, L.J.~King, G.-B.~Zhao and H.~Zhao, \emph{An analytic ray-tracing
  algorithm for weak lensing},
  \href{https://doi.org/10.1111/j.1365-2966.2011.18754.x}{\emph{Monthly Notices
  of the Royal Astronomical Society} {\bfseries 415} (2011) 881}
  [\href{https://arxiv.org/abs/1012.1625}{{\ttfamily 1012.1625}}].

\bibitem{barreira_ray-ramses_2016}
A.~Barreira, C.~Llinares, S.~Bose and B.~Li, \emph{{RAY}-{RAMSES}: a code for
  ray tracing on the fly in {N}-body simulations},
  \href{https://doi.org/10.1088/1475-7516/2016/05/001}{\emph{Journal of
  Cosmology and Astroparticle Physics} {\bfseries 2016} (2016) 001}.

\bibitem{KilledarEtAl2012}
M.~Killedar, P.D.~Lasky, G.F.~Lewis and C.J.~Fluke, \emph{Gravitational lensing
  with three-dimensional ray tracing},
  \href{https://doi.org/10.1111/j.1365-2966.2011.20023.x}{\emph{Monthly Notices
  of the Royal Astronomical Society} {\bfseries 420} (2012) 155}.

\bibitem{bar-kana_gravitational_1997}
R.~Bar-Kana, \emph{Gravitational lensing as a probe of dark matter, the
  distance scale, and gravitational waves in the universe}, Ph.D. thesis, May,
  1997.

\bibitem{quinn_time_1997}
T.~Quinn, N.~Katz, J.~Stadel and G.~Lake, \emph{Time stepping {N}-body
  simulations},  Oct., 1997.

\bibitem{saha_symplectic_1992}
P.~Saha and S.~Tremaine, \emph{Symplectic {Integrators} for {Solar} {System}
  {Dynamics}}, \href{https://doi.org/10.1086/116347}{\emph{The Astronomical
  Journal} {\bfseries 104} (1992) 1633}.

\bibitem{springel_simulating_2021}
V.~Springel, R.~Pakmor, O.~Zier and M.~Reinecke, \emph{Simulating cosmic
  structure formation with the {\textless}span
  style="font-variant:small-caps;"{\textgreater}gadget{\textless}/span{\textgreater}
  -4 code}, \href{https://doi.org/10.1093/mnras/stab1855}{\emph{Monthly Notices
  of the Royal Astronomical Society} {\bfseries 506} (2021) 2871}.

\bibitem{yoshida_recent_1993}
H.~Yoshida, \emph{Recent progress in the theory and application of symplectic
  integrators}, \href{https://doi.org/10.1007/BF00699717}{\emph{Celestial
  Mechanics and Dynamical Astronomy} {\bfseries 56} (1993) 27}.

\bibitem{dodelson_modern_2003}
S.~Dodelson, \emph{Modern cosmology}, Academic Press, San Diego, Calif (2003).

\bibitem{schneider_gravitational_1992}
P.~Schneider, J.~Ehlers and E.E.~Falco, \emph{Gravitational {Lenses}},
  Astronomy and {Astrophysics} {Library}, Springer Berlin Heidelberg, Berlin,
  Heidelberg (1992),
  \href{https://doi.org/10.1007/978-3-662-03758-4}{10.1007/978-3-662-03758-4}.

\bibitem{hockney_computer_2021}
R.~Hockney and J.~Eastwood, \emph{Computer {Simulation} {Using} {Particles}},
  CRC Press, 0~ed. (Mar., 2021),
  \href{https://doi.org/10.1201/9780367806934}{10.1201/9780367806934}.

\bibitem{epstein_post-post-newtonian_1980}
R.~Epstein and I.I.~Shapiro, \emph{Post-post-{Newtonian} deflection of light by
  the {Sun}}, \href{https://doi.org/10.1103/PhysRevD.22.2947}{\emph{Physical
  Review D} {\bfseries 22} (1980) 2947}.

\bibitem{planck_collaboration_planck_2016}
{Planck Collaboration}, P.A.R.~Ade, N.~Aghanim, M.~Arnaud, M.~Ashdown,
  J.~Aumont et~al., \emph{\textit{{Planck}} 2015 results: {XIII}.
  {Cosmological} parameters},
  \href{https://doi.org/10.1051/0004-6361/201525830}{\emph{Astronomy \&
  Astrophysics} {\bfseries 594} (2016) A13}.

\bibitem{naiman_first_2018}
J.P.~Naiman, A.~Pillepich, V.~Springel, E.~Ramirez-Ruiz, P.~Torrey,
  M.~Vogelsberger et~al., \emph{First results from the {IllustrisTNG}
  simulations: {A} tale of two elements -- chemical evolution of magnesium and
  europium}, \href{https://doi.org/10.1093/mnras/sty618}{\emph{Monthly Notices
  of the Royal Astronomical Society} {\bfseries 477} (2018) 1206}.

\bibitem{springel_first_2018}
V.~Springel, R.~Pakmor, A.~Pillepich, R.~Weinberger, D.~Nelson, L.~Hernquist
  et~al., \emph{First results from the {IllustrisTNG} simulations: matter and
  galaxy clustering},
  \href{https://doi.org/10.1093/mnras/stx3304}{\emph{Monthly Notices of the
  Royal Astronomical Society} {\bfseries 475} (2018) 676}.

\bibitem{marinacci_first_2018}
F.~Marinacci, M.~Vogelsberger, R.~Pakmor, P.~Torrey, V.~Springel, L.~Hernquist
  et~al., \emph{First results from the {IllustrisTNG} simulations: radio haloes
  and magnetic fields},
  \href{https://doi.org/10.1093/mnras/sty2206}{\emph{Monthly Notices of the
  Royal Astronomical Society} (2018) }.

\bibitem{nelson_first_2018}
D.~Nelson, A.~Pillepich, V.~Springel, R.~Weinberger, L.~Hernquist, R.~Pakmor
  et~al., \emph{First results from the {IllustrisTNG} simulations: the galaxy
  color bimodality}, \href{https://doi.org/10.1093/mnras/stx3040}{\emph{Monthly
  Notices of the Royal Astronomical Society} {\bfseries 475} (2018) 624}.

\bibitem{pillepich_first_2018}
A.~Pillepich, D.~Nelson, L.~Hernquist, V.~Springel, R.~Pakmor, P.~Torrey
  et~al., \emph{First results from the {IllustrisTNG} simulations: the stellar
  mass content of groups and clusters of galaxies},
  \href{https://doi.org/10.1093/mnras/stx3112}{\emph{Monthly Notices of the
  Royal Astronomical Society} {\bfseries 475} (2018) 648}.

\bibitem{petri_sample_2016}
A.~Petri, Z.~Haiman and M.~May, \emph{Sample variance in weak lensing: how many
  simulations are required?},
  \href{https://doi.org/10.1103/PhysRevD.93.063524}{\emph{Physical Review D}
  {\bfseries 93} (2016) 063524}.

\bibitem{ZhangLiEtAl}
Y.~Zhang, Y.~Li, D.~Jamieson, L.~Lu and et~al., \emph{Neural and symbolic
  optimization of cosmological particle-mesh simulation}, {\emph{in prep} }.

\bibitem{takahashi_revising_2012}
R.~Takahashi, M.~Sato, T.~Nishimichi, A.~Taruya and M.~Oguri, \emph{Revising
  the {Halofit} {Model} for the {Nonlinear} {Matter} {Power} {Spectrum}},
  \href{https://doi.org/10.1088/0004-637X/761/2/152}{\emph{The Astrophysical
  Journal} {\bfseries 761} (2012) 152}.

\bibitem{Hu_2022}
L.~Hu, A.~Heavens and D.~Bacon, \emph{Light bending by the cosmological
  constant}, {\emph{Journal of Cosmology and Astroparticle Physics} {\bfseries
  2022} (2022) 009}.

\bibitem{dalal_hyper_2023}
R.~Dalal, X.~Li, A.~Nicola, J.~Zuntz, M.A.~Strauss, S.~Sugiyama et~al.,
  \emph{Hyper {Suprime}-{Cam} {Year} 3 {Results}: {Cosmology} from {Cosmic}
  {Shear} {Power} {Spectra}},  Apr., 2023.

\bibitem{Schneider1985}
P.~Schneider, \emph{A new formulation of gravitational lens theory, time-delay,
  and {{Fermat}}'s principle}, {\emph{Astronomy and Astrophysics} {\bfseries
  143} (1985) 413}.

\bibitem{BlandfordNarayan1986}
R.~Blandford and R.~Narayan, \emph{Fermat's principle, caustics, and the
  classification of gravitational lens images},
  \href{https://doi.org/10.1086/164709}{\emph{The Astrophysical Journal}
  {\bfseries 310} (1986) 568}.

\end{thebibliography}\endgroup

\appendix
\section{Hamiltonian dynamics of photons lensed by weak gravitational field}
\label{app:hamiltonian}

In this section, we derive the EOMs in \crefrange{eqn:eom1}{eqn:eom3}
from the Hamiltonian principle.
Although the procedure is similar to that of
Ref.~\cite{bar-kana_gravitational_1997}, we employ a different metric
and present the derivations in full detail.
These details, not included in Ref.~\cite{bar-kana_gravitational_1997},
may prove helpful to some readers.

The general relativistic covariant Hamiltonian for a photon is given by
\cref{eqn:hamiltonian_3d} and copied here:
\begin{equation}
    H(x^j,p_i,\tau) = c \sqrt{h^{ij}p_i p_j}(1+2\frac{\npot}{c^2}) \,.
\end{equation}
The photon's EOMs are given by Hamilton's equations
\begin{align}
    \label{eqn:hamiltons_equations1}
    \frac{dx^i}{d\tau} & = \{ x^i, H\}
    = \frac{\partial H}{\partial p_i} \,, \\
    \label{eqn:hamiltons_equations2}
    \frac{dp_i}{d\tau} & =  \{ p_i, H\}
    = -\frac{\partial H}{\partial x^i} \,,
\end{align}
which we will solve explicitly.
The time derivatives of position and momentum are
\begin{align}
\frac1c \frac{dx^j}{d\tau}
&= \frac{\partial \sqrt{h^{kl}p_k p_l}}{\partial p_j}
  \Bigl(1 + 2\frac\npot{c^2}\Bigr)
= \frac{h^{kl}}{2p} \Bigl( \frac{\partial p_k}{\partial p_j} p_l
    + \frac{\partial p_l}{\partial p_j} p_k
  \Bigr) \Bigl(1 + 2\frac\npot{c^2}\Bigr)
= n^j \Bigl(1 + 2\frac\npot{c^2}\Bigr) \,. \\
\frac1c \frac{dp_j}{d\tau}
&= -2p \frac{\partial_j \npot}{c^2}
  - \frac{p_k p_l \partial_j h^{kl}}{2p}
    \Bigl(1 + 2\frac\npot{c^2}\Bigr)
= - p \Bigl[ 2 \frac{\partial_j \npot}{c^2}
    + \frac{ n_k n_l \partial_j h^{kl}}{2}
      \Bigl(1 + 2\frac\npot{c^2}\Bigr) \Bigr] \,,
\end{align}
where the unit momentum vector $n^i$ is
\begin{equation}
n^i \equiv \frac{p^i}p = \frac{p^i}{\sqrt{h^{ij}p_i p_j}} \,.
\end{equation}

$n^i$ is the most important dynamical variable in ray-tracing.
To explicitly derive its EOM, we express its time dependence in those of
the two independent variables $p_i$ and $x^j$, and consider the
following expansion
\begin{equation}
\frac1c \frac{dn^i}{d\tau}
= \frac1c \frac{d}{d\tau} \Bigl( \frac{p^i}p \Bigr)
= \frac{\partial (p^i/p)}{\partial p_j} \frac1c \frac{d p_j}{d \tau}
  + \frac{\partial (p^i/p)}{\partial x^j} \frac1c \frac{d x^j}{d \tau} \,.
\end{equation}
The first term becomes
\begin{align}
\nonumber
\frac{\partial (p^i/p)}{\partial p_j} \frac1c \frac{d p_j}{d \tau}
&= \frac1c \frac{d p_j}{d \tau} \Bigl(
  \frac{h^{ik}}{p} \frac{\partial p_k}{\partial p_j}
  - \frac{p^i}{p^2} \frac{\partial p}{\partial p_j} \Bigr)
= \frac1c \frac{d p_j}{d \tau} \frac{h^{ij} - n^i n^j}{p} \\
&= \Bigl[ -2 \frac{\partial_j \npot}{c^2}
  - \frac{n_k n_l \partial_j h^{kl}}2 \Bigl(1 + 2\frac\npot{c^2}\Bigr)
\Bigr] (h^{ij} - n^i n^j) \,.
\end{align}
Meanwhile, the second term becomes
\begin{align}
\nonumber
\frac{\partial (p^i/p)}{\partial x^j} \frac1c \frac{d x^j}{d \tau}
&= \frac1c \frac{d x^j}{d \tau} \Bigl( \frac{p_k}{p} \partial_j h^{ik}
  - \frac{p^i}{p^2} \partial_j p \Bigr)
= \frac1c \frac{d x^j}{d \tau} \Bigl( n_k \partial_j h^{ik}
  - \frac12 n^i n_k n_l \partial_j h^{kl} \Bigr) \\
&= \Bigl( n^j n_k \partial_j h^{ik}
  - \frac12 n^i n^j n_k n_l \partial_j h^{kl} \Bigr)
\Bigl(1 + 2\frac\npot{c^2}\Bigr) \,.
\end{align}
Putting the two terms together,
\begin{align}
\nonumber
\frac1c \frac{dn^i}{d\tau}
=& -2 \frac{\partial_j \npot}{c^2} (h^{ij} - n^i n^j)
  - \frac{n_k n_l \partial_j h^{kl}}2 \Bigl(1 + 2\frac\npot{c^2}\Bigr)
    (h^{ij} - \cancel{n^i n^j}) \\
\nonumber
&+ \Bigl( n^j n_k \partial_j h^{ik}
  - \cancel{\frac12 n^i n^j n_k n_l \partial_j h^{kl}} \Bigr)
\Bigl(1 + 2\frac\npot{c^2}\Bigr) \\
=& -2 \frac{\partial_j \npot}{c^2} (h^{ij} - n^i n^j)
  - \frac{h^{ij} n_k n_l \partial_j h^{kl}}2
    \Bigl(1 + 2\frac\npot{c^2}\Bigr)
  + n^j n_k \partial_j h^{ik} \Bigl(1 + 2\frac\npot{c^2}\Bigr) \,.
\end{align}
It turns out that we can simplify the last two terms, because
\begin{align}
\nonumber
- \frac12 h^{ij} n_k n_l \partial_j h^{kl} + n^j n_k \partial_j h^{ik}
&= \frac12 h^{ij} n^k n^l \partial_j h_{kl}
  - h^{il} n^j n^k \partial_j h_{kl} \\
\nonumber
&= \frac12 h^{il} n^j n^k (\partial_l h_{jk} - 2 \partial_j h_{kl}) \\
\nonumber
&= - \frac12 h^{il} n^{(j} n^{k)}
  (\partial_j h_{kl} + \partial_k h_{jl} - \partial_l h_{jk}) \\
&= - \Gamma^i_{jk} n^j n^k \,,
\end{align}
where we have used $\partial_j h^{ik} = - h^{il} h^{km} \partial_j
h_{lm}$ in the first equality, swapped $j$ and $l$ in the (first term of
the) second one, symmetrized $j$ and $k$ in the third, and reduced the
metric derivatives to the Christoffel symbol $\Gamma^i_{jk}$ at last.
With the result above, we can simplify $dn^i / d\tau$ as follows
\begin{equation}
\frac1c \frac{dn^i}{d\tau}
= -2 \frac{\partial_j \npot}{c^2} (h^{ij} - n^i n^j)
  - \Gamma^i_{jk} n^j n^k \Bigl(1 + 2\frac\npot{c^2}\Bigr) \,,
\end{equation}
which agrees with the result in \cite{bar-kana_gravitational_1997}.

We are now ready to derive the EOMs of a photon.
Until this point, we have not enforced any specific metric on the EOMs.
We now select the metric in \cref{eqn:metric_3d} with coordinates $\bfx
= (\bftheta, \chi)$.
Here, we assume small angle approximations such that $\sin^2(\theta)
\approx 1$.
The unit momentum vector in this coordinate system is then
\begin{equation}
\bfn = \Bigl[ \frac{\bfv}{cr},
  -1 + \order\Bigl(\frac{\bfv^2}{c^2}\Bigr) \Bigr] \,,
\end{equation}
where the first term can be viewed as the normalized peculiar angular
velocity on the sky, with $\bfv$ being the 2D transverse peculiar
velocity.
We shall work in the limits of weak fields and small angles, which
corresponds to first order in $\npot/c^2$, $\grad \npot/c^2$, and
$\bfv/c$.
The nonzero Christoffel symbols are
\begin{equation}
\Gamma^1_{13} = \Gamma^1_{31} = \Gamma^2_{23} = \Gamma^2_{32}
  = \frac{r'}{r} \,, \qquad
\Gamma^3_{11} = \Gamma^3_{22} = - r r' \,,
\end{equation}
where $'$ denotes derivative with respect to $\chi$.
We have omitted corrections of $\order(\grad \npot / c^2)$ because the
Christoffel symbol only appears in the expression $\Gamma^i_{jk} n^j n^k
(1 + 2\npot / c^2)$, and always pairs with at least one $\order(v/c)$
term.
Taking time derivatives of $n^i$ for $i = 1, 2$, we obtain
\begin{equation}
\frac1c \frac{dn^i}{d\tau}
= \frac1{c^2 r}\frac{dv^i}{d\tau} - \frac{v^i}{c^2 r^2}\frac{dr}{d\tau}
\end{equation}
Consider $i=1$ of the $n^i$ EOM:
\begin{align}
\nonumber
\frac1c \frac{dn^1}{d\tau}
&= \frac{1}{c^2} (
    - 2 h^{11} \partial_1 \npot + 2 n^1 n^1 \partial_1 \npot
    + 2 n^1 n^2 \partial_2 \npot + 2 n^1 n^3 \partial_3 \npot)
  - 2\Gamma^1_{13} n^1 n^3 \Bigl(1 + 2\frac{\npot}{c^2}\Bigr) \\
&\simeq -2 \frac{\partial_1 \npot}{c^2 r^2}
  + 2\frac{v^1 r'}{c r^2} \,.
\end{align}
The general case follows similarly by symmetry:
\begin{equation}
\frac1c \frac{dn^i}{d\tau} = -2 \frac{\partial_i \npot}{c^2 r^2}
+ 2\frac{v^i}{c} \frac{r'}{r^2} \,,
\end{equation}
where we ignore the potential and its gradient terms since they are
higher order corrections (while more generally $h^{ij} - n^i n^j$
projects in the transverse direction of $n^i$).
Comparing this with the above equation, and approximating $c d\tau
\approx - d\chi$ by ignoring time delay correction of $\order(\npot /
c^2)$ and $\order(v^2 / c^2)$, we derive
\begin{equation}
\frac{d\eta^i} {d\chi}
\equiv \frac{d(r v^i / c)} {d\chi}
= \frac{r v^{i\prime} + r' v^i}c
= 2 \frac{\partial_i\npot}{c^2}
= 2 \frac1{c^2} \frac{\partial\npot}{\partial \theta^i} \,.
\end{equation}

\section{Symplectic integrator}
\label{app:symp}

We have purposely written the EOMs in \cref{eqn:eom1,eqn:eom2} using
$\bftheta$ and $\bfeta$, in unit of angle and length, respectively.
Either from the Hamiltonian \cref{eqn:hamiltonian_3d} or from the EOMs,
we see that the system, under the assumptions above, admits a separable
Hamiltonian
\begin{equation}
\label{eqn:H_tilde}
\tilde H(\bftheta, \bfeta, \chi)
= \underbrace{\frac{\bfeta^2}{2r^2}}_{T}
  + \underbrace{2\frac{\npot(\bftheta, \chi)}{c^2}}_{V} \,,
\end{equation}
where $\bftheta$ and $\bfeta$ are the canonical coordinate and momentum,
$\chi$ serves the function of time, and $T$ and $V$ represent kinetic
and potential energy.
In fact, $\tilde H$ also follows from the Fermat principle, and can be
derived with Legendre transformation from the Lagrangian in the Fermat
action (see \cite{bar-kana_gravitational_1997, Schneider1985,
BlandfordNarayan1986}, especially for the continuous version in the
former).
This allows us to decouple the effect of the kinetic and the potential
energy and rewrite the EOMs in \cref{eqn:eom1,eqn:eom2} as
\begin{equation}
\label{eqn:eom_poisson}
\frac{d\bftheta}{d\chi} = \{\bftheta, T\} \,, \qquad
\frac{d\bfeta}{d\chi} = \{\bfeta, V\} \,.
\end{equation}
where $\{\}$ is the Poisson bracket.

Formally, we can integrate\footnote{Here, for an operator $A$ and a
vector $\bfv$, we use the shorthand $A\bfv = \{\bfv, A\}$.}
\cref{eqn:eom_poisson} over the interval $[\chi, \chi+ \Delta \chi]$ by
\begin{equation}
    \begin{bmatrix}
           \bftheta \\
           \bfeta \\
    \end{bmatrix}
    (\chi+\Delta\chi)
    = e^{(T+V) \Delta \chi}
    \begin{bmatrix}
           \bftheta \\
           \bfeta \\
    \end{bmatrix}(\chi) \,.
\end{equation}
Since the EOMs of $\bftheta$ and $\bfeta$ are coupled together, we
cannot solve them individually.
The KDK integrator utilizes the Baker-Campbell-Hausdorff identity to
approximate this evolution operator as the product
\begin{equation}
\label{eqn:BCH}
    e^{(H + \tilde H_\mathrm{err}) \Delta \chi}
    =
    e^{V \frac{\Delta \chi}{2}}
    e^{T {\Delta \chi}}
    e^{V \frac{\Delta \chi}{2}}
\end{equation}
where $\tilde H_{\mathrm{err}}$ denotes the approximation error, which
is second order in $\Delta \chi$ \cite{yoshida_recent_1993}.
\Cref{eqn:BCH} shows that we can decompose each $\Delta \chi$
integration at each time step into three consecutive steps, each
updating only either the position or the momentum variables.
$e^{V \Delta \chi}$ is termed a kick operator since it updates $\bfeta$
and leaves $\bftheta$ unchanged, while $e^{T {\Delta \chi}}$ is called a
drift operator since it updates $\bftheta$ and leaves $\bfeta$
unchanged.

\section{Adaptive ray mesh}
\label{app:ray_mesh}

The 2D array of light rays is characterized by its pixel size
$\pixsize$, i.e.\ their spacing at $z=0$, and the number of pixels
$M_{x, y}$.
In order to interpolate and transfer the PM forces from the 3D mesh to the
rays, we need another intermediate 2D angular mesh of resolution
$\restwod$ and size $N_{x, y}$.
We call it the ray mesh (and likewise the 3D mesh particle mesh), as
explained in \cref{tab:variables}.

Generally, $\restwod \leq \pixsize$ and $N_{x,y} > M_{x,y}$.
Our adaptive ray mesh needs to address two problems: 1) we need to vary
the 2D mesh resolution $\restwod$ as a function of $\chi$
(\cref{sec:ray_mesh}) and 2) we need to add padding to the ray mesh.
Padding is crucial for accurate smoothing and deconvolution in the
computation of the kick operator, as it alleviates the effects of
periodic boundary assumptions inherent to FFT.

A 3D mesh cell at $\chi$ corresponds to an angular size $\resthreed /
r(\chi)$.
Therefore, for a lens plane that spans $[\chi_a, \chi_b]$, we define the
angular resolution limit $\reslim$ as in \cref{eqn:reslim}:
\begin{equation}
\reslim = \max\Bigl(
  \frac{2 \, \resthreed}{r_a + r_b},\pixsize \Bigr) \,,
\end{equation}
in which the former dominates near the observer as limited by the PM
force resolution, and the latter takes effect at early times.
Let us fix the ray mesh spacing as
\begin{equation}
    \restwod = \iota \reslim \,,
\end{equation}
where $0 < \iota \leq 1$ is an accuracy parameter.
The smaller $\iota$ is, the finer the ray mesh.
Take the x-direction for example, we want
\begin{equation}
N_x \restwod \geq M_x \pixsize + p_{\min}\lambda_{\lim} \,,
\end{equation}
where $p_{\min}$ is the minimum padding width in unit of $\reslim$.
The above conditions reduce to
\begin{equation}
N_x = \Bigl\lceil M_x \frac{\pixsize}{\restwod}
  + \frac{p_{\min}}{\iota} \Bigr\rceil_\mathrm{FFT} \,.
\end{equation}
where we round up the mesh size to an integer that is efficient for FFT,
e.g.\ powers of 2, while also saving the number of compilations that can
affect \jax\ performance.
The y-dimension follows accordingly.

\section{Other considerations and convergence tests for point mass lensing}
\label{sec:point_mass_error}

\begin{figure}[tbh]
\centering
\includegraphics[width=0.6\textwidth]{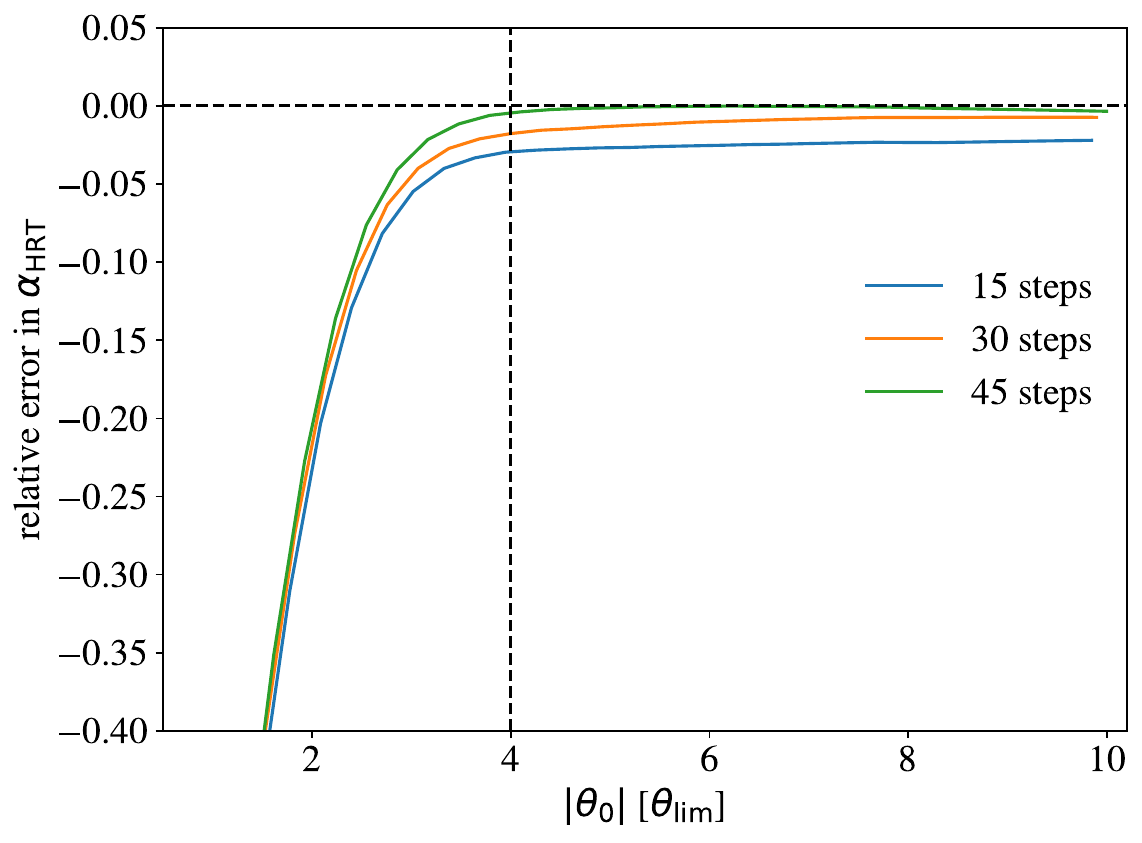}
\caption{Relative error of $\alphahrt$ versus $\thetathresh$ as a
function of the number of HRT integration steps used for a point mass
lensing task.
This figure is to be compared to the blue line in
\cref{fig:pointmass_resolution_mass} which has the same experimental
setup with $45$ integration steps.}
\label{fig:pointmass_timestep}
\end{figure}

In the context of point mass lensing, we explore how the accuracy of HRT
is influenced by the boundary conditions of the 3D mesh, the number of
integration steps, the 2D ray mesh accuracy parameter $\iota$, and the
2D ray mesh padding size $p_{\min}$ (\cref{app:ray_mesh}).

First, when we perform ray-tracing on a point mass lens, we compute the
gravitational potential field using FFT with periodic boundary
conditions.
This is an approximation of the gravity solver and not the ray-tracing
algorithm itself.
To disentangle the error induced by the gravity solver from those due to
HRT, we have compared $\alphahrt$ against a theoretical result that
incorporates periodic boundary conditions throughout this work.
To model the effect of the periodic boundary condition, we assume there
is not only one but also infinitely many periodic images of the point
mass lens, at comoving positions $\bfm \in \{ (0, 0), (\pm L_x, 0), (0,
\pm L_y), \cdots \}$, where $L_{x,y}$ are the side
lengths of the simulation box.
The leading order deflection in \cref{eqn:pointmass_analytic} is
replaced by its periodic summation:
\begin{align}
\label{eqn:pointmass_periodic}
\hat\alpha \frac{\bftheta_0}{\left|\bftheta_0\right|}
= \frac{4GM}{c^2} \frac{\bfb}{b^2}
\longrightarrow \frac{4GM}{c^2}
  \sum_\bfm \frac{\bfb-\bfm}{\lvert\bfb-\bfm\rvert^2} \,,
\end{align}
where $\bfb = (\theta_{0x} \chi_\mathrm{l}, ~\theta_{0y}
\chi_\mathrm{l})$ is the impact parameter of the photons.
\Cref{eqn:pointmass_periodic} reduces to \cref{eqn:pointmass_analytic}
when only $\bfm = (0, 0)$ is considered.
Throughout this work, we account for the nearest $32^2$ periodic images
of the lens mass when computing $\alphatheory$.
The theoretical result converges within numerical precision.

\begin{figure}[tbh]
\centering
\includegraphics[width=0.6\textwidth]{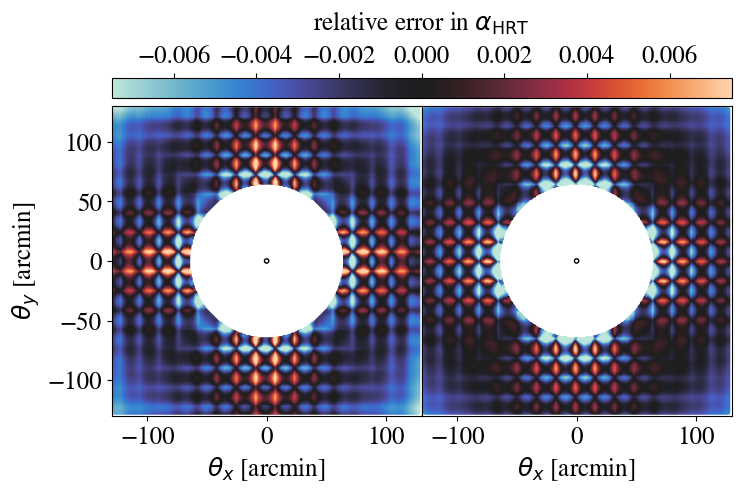}
\caption{The effect of ray mesh hyper-parameters: the accuracy parameter
$\iota$ and the padding size $p_{\min}$.
The two panels show $\alphahrt$'s residual errors on the image plane for
the point mass test.
The left panel uses $(\iota = 1, ~ p_{\min} = 4)$ and the right panel
uses $(\iota = 1, ~ p_{\min} = 256)$.
They are compared to panel (b) of \cref{fig:pointmass_resolution} which
uses $(\iota = 0.5, ~ p_{\min} = 256)$.}
\label{fig:pointmass_padding}
\end{figure}

The numerical experiments in \cref{sec:pointmass} use 45 integration
steps and assume a padding of $\iota = 0.5$ and $p_{\min} = 256$.
We utilize the same setup as in \cref{sec:pointmass} to evaluate how the
accuracy of HRT depends on these hyper-parameters.
\Cref{fig:pointmass_timestep} illustrates that $\alphahrt$ converges to
the theoretical values as the number of integration steps increases.
With too few integration steps, $\alphahrt$ is typically lower than the
truth.
The two panels in \cref{fig:pointmass_padding} demonstrate the effect of
the ray mesh hyper-parameters: the accuracy parameter $\iota$ and the
padding size $p_{\min}$.
The left panel uses $(\iota = 1, ~ p_{\min} = 4)$ and the right panel
uses $(\iota = 1, ~ p_{\min} = 256)$.
These are compared to panel (b) of \cref{fig:pointmass_resolution},
which has the same experimental setup but with $(\iota = 0.5, ~ p_{\min}
= 256)$.
In both cases, HRT achieves sub-percent accuracy, showing that increased
padding improves $\alphahrt$ accuracy near the boundary and higher
$\iota$ results in more prominent discretization features of the
particle and ray meshes.
The result also shows that, in practice, we do not need to carry a
$p_{\min}$ that is as large.

\section{Convergence tests for weak lensing power spectrum}
\label{app:cl}

\begin{figure}[tbh]
\centering
\includegraphics[width=0.49\textwidth]{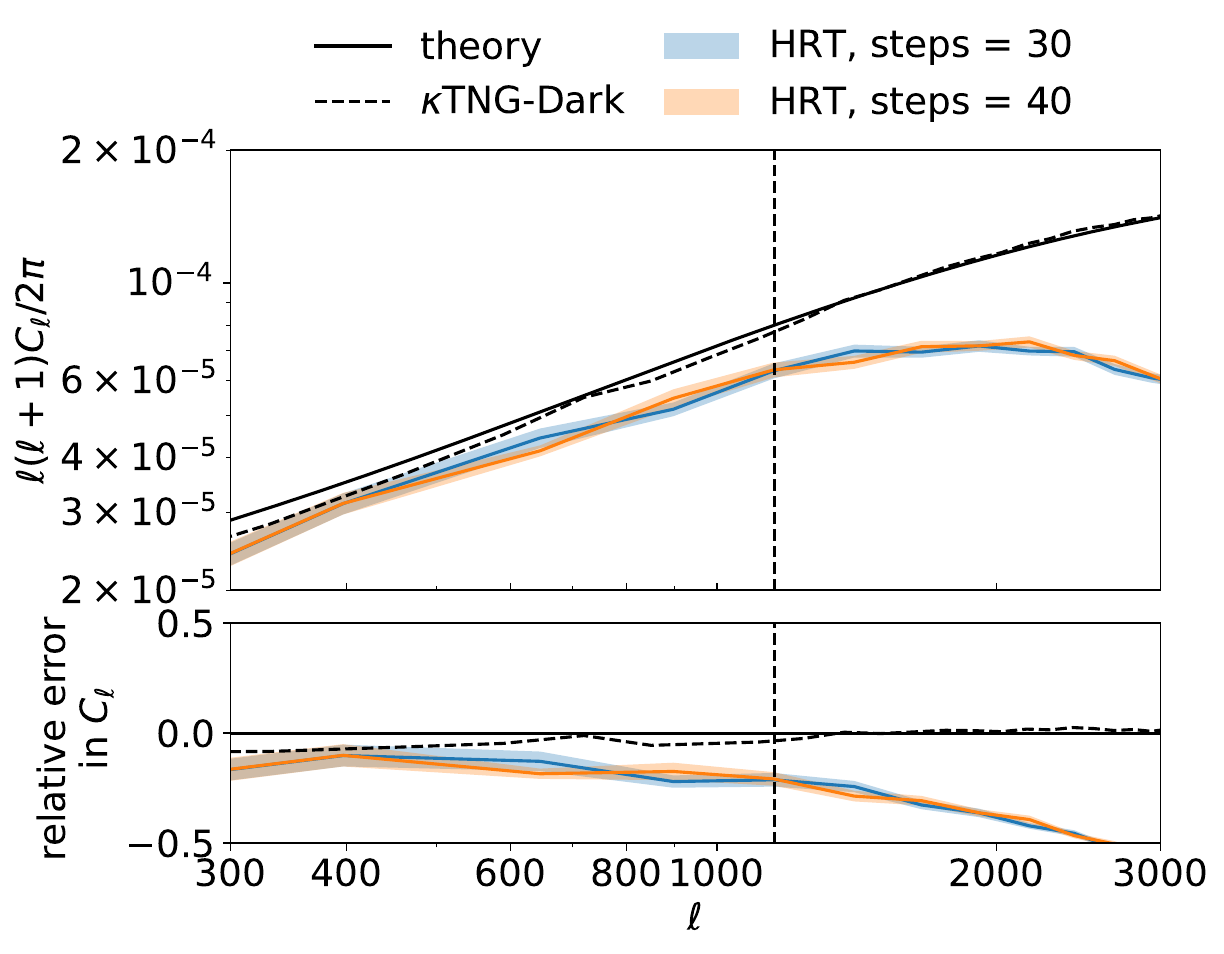}
\includegraphics[width=0.49\textwidth]{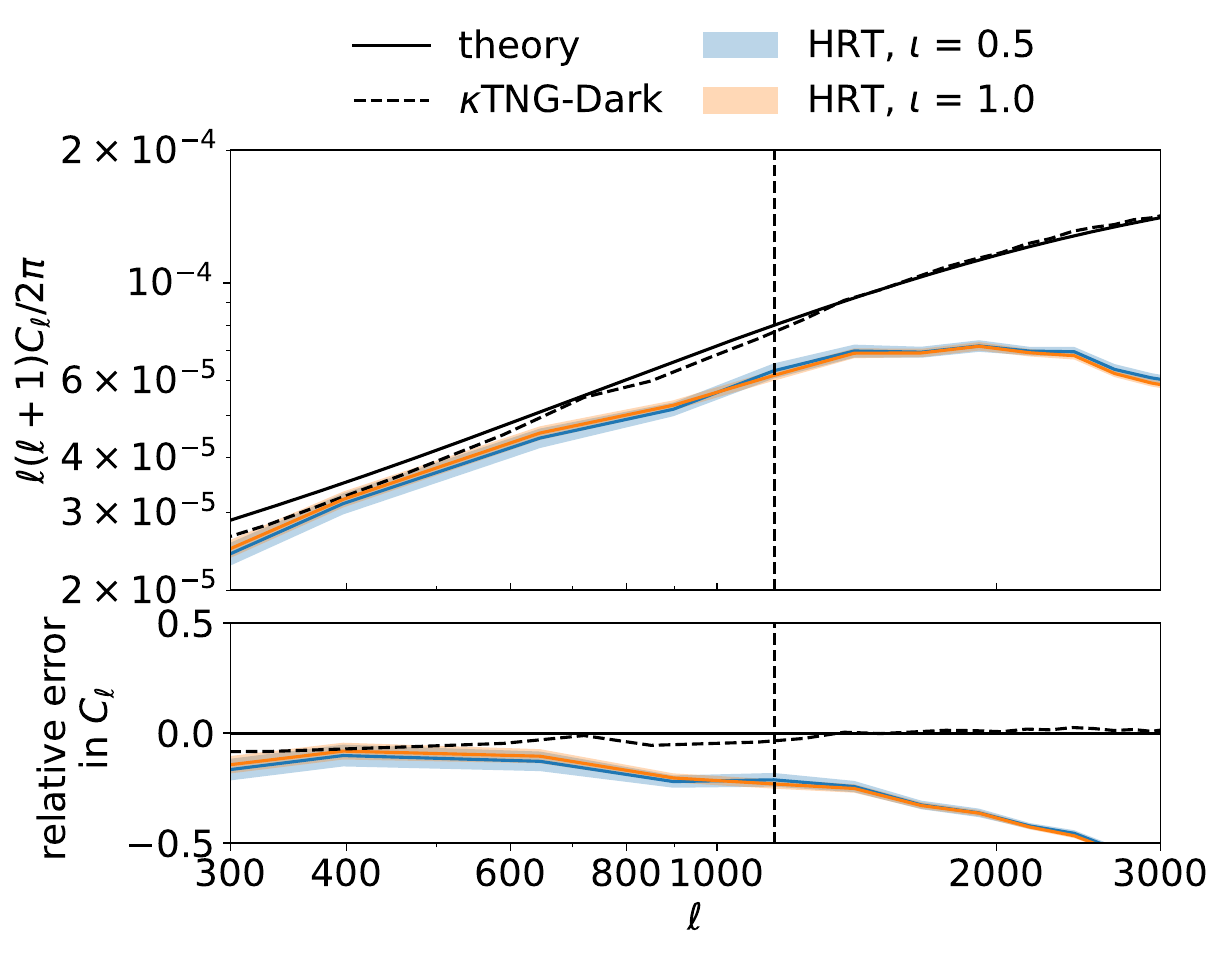}
\caption{\response{Relative error of the SO-enhanced $C_\ell$ as a function of the number of HRT
integration steps (left) and of the HRT ray mesh accuracy parameter
$\iota$ (right). The shaded regions show the standard error of the mean over 25 independent simulations. Similar to \cref{fig:cl}, the black solid line shows the theory prediction, the dashed line shows the $\ktng$ simulation, and the vertical dotted line shows the $\ell = 1200$ scale cut.}}
\label{fig:cl_tests}
\end{figure}

\response{We compute the convergence power spectrum by ray tracing to a source
plane at $z_\mathrm{s} = 1.034$.
We tile the past light cone with simulation boxes of a $512^3$ mesh and
$512^3$ particles with a particle/mesh spacing of $0.4~\Mpch$}, similar
to \cref{sec:cl}.
We vary the number of ray tracing time steps and the ray mesh accuracy
parameter $\iota$, as shown in \cref{fig:cl_tests}, and find that the
power spectrum is not sensitive to these hyper-parameters.

\end{document}